\documentclass[journal]{IEEEtran}
\pdfoutput=1
\usepackage{comment}
\usepackage{amssymb}
\usepackage{multirow}
\usepackage{tikz}
\usepackage{topcapt}
\usepackage{tablefootnote}
\usepackage{color,colortbl}
\definecolor{grey}{gray}{0.6}
\newcommand{\tabincell}[2]{\begin{tabular}{@{}#1@{}}#2\end{tabular}}
\usepackage{ifpdf}
\usepackage{cite}

\usepackage[cmex10]{amsmath}
\usepackage{url}
\usepackage{graphicx}

\gdef\PHEV{$^{\diamond}$}
\usetikzlibrary{decorations.pathmorphing} 
\usetikzlibrary{fit}					
\usetikzlibrary{backgrounds}	
\title{Charging Electric Vehicles in the Smart City: A Survey of Economy-driven Approaches}
\author{Wenjing~Shuai, Patrick~Maill\'e, Alexander~Pelov
\thanks{The three authors are with the Networks, Security, Multimedia Department, Institut Mines-Telecom/Telecom Bretagne, 2 rue de la chataigneraie CS 17607, 35576 Cesson-S\'evign\'e, FRANCE e-mail: \{first\}.\{last\}@telecom-bretagne.eu}%
\thanks{Manuscript received.}
}

\newtheorem{example}{Example}

\begin{document}

\maketitle

\begin{abstract}
Electric Vehicles (EVs), as their penetration increases, are not only challenging the sustainability of the power grid, but also stimulating and promoting its upgrading. Indeed, EVs can actively reinforce the development of the Smart Grid if their charging processes are properly coordinated through two-way communications, possibly benefiting all types of actors. 

Because grid systems involve a large number of actors with nonaligned objectives, we focus on the economic and incentive aspects, where each actor behaves in its own interest. We indeed believe that the market structure will directly impact the actors' behaviors, and as a result the total benefits that the presence of EVs can earn the society, hence the need for a careful design. 
This survey provides an overview of economic models considering unidirectional energy flows, but also bidirectional energy flows, i.e., with EVs temporarily providing energy to the grid. We describe and compare the main approaches, summarize the requirements on the supporting communication systems, and propose a classification to highlight the most important results and lacks.
\end{abstract}

%
%
\section{Introduction}
\label{sec:Intro}
\IEEEPARstart{D}{iminishing} oil supply and increasing environmental concerns strongly motivate research efforts toward the electrification of transportation, and technological advances have fostered a rapid arrival of Electric Vehicles (EVs) in the market. 
However, the charging of EVs has a tremendous impact on the stakeholders in both the electricity and transportation domains, such as electricity producers, power grid operators, policy makers, retailers, and customers~\cite{Dickerman2010IntroEV}. The EV load can drive electricity prices up~\cite{Balram2012Impact}, and alter the producers' generation portfolios, resulting in an increase of CO$_2$ emission~\cite{Foley2013Impact}. 
Additionally, high penetration with uncontrolled charging threatens the sustainability of distribution networks~\cite{Green2011ImpactReview,Shafiee2013Impact}. 
For example, for an EV penetration of 25\%, almost 30\% of network facilities would need to be upgraded, while this ratio drops to 5\% if the charging load can be shifted to less crowded time periods~\cite{Dow2010Impact}. These research works reach a consensus that EV charging should be controlled to avoid distribution congestion and higher peak-to-average ratios (i.e., demand sporadicity). 

\begin{figure}[htbp]
   \centering
   \includegraphics[width=8cm]{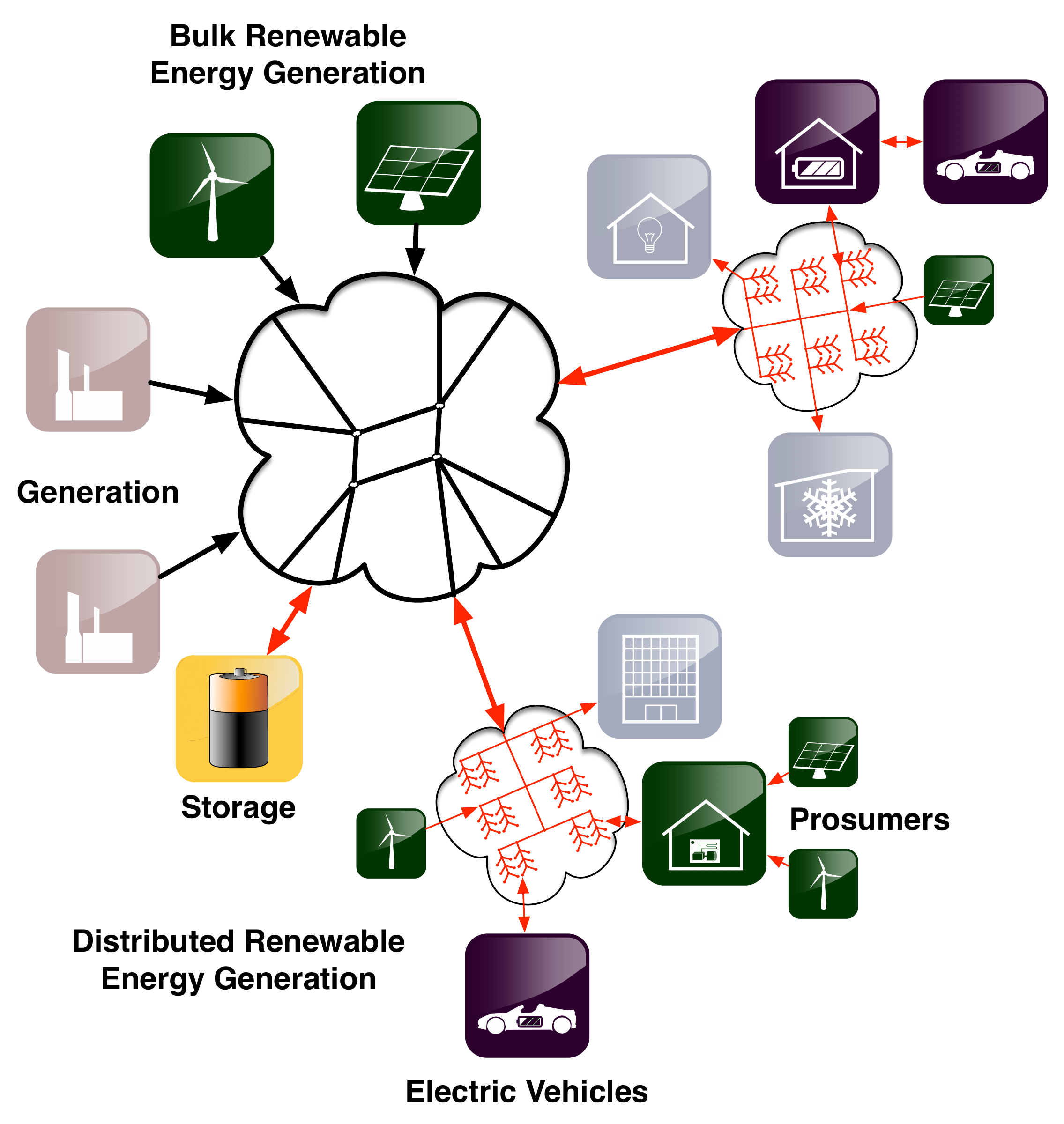} 
   \centering
   \caption{Actors and energy flows in the Smart Grid}
   \label{fig:SmartGrid}
\end{figure}

At the same time, the Power Grid is witnessing one of its major evolutions since its conception at the beginning of the past century. 
The classical structure of electricity being produced in a small number of big, centralized, power plants, and flowing through the transmission and distribution networks to be consumed by end users is being challenged by the increasing penetration of renewable energy sources. 
The possibility to communicate bidirectionally with all elements of the grid--and as a consequence to achieve unprecedented levels of monitoring and control--serves as a major technological enabler of the new Smart Grid, allowing to accommodate new types of demand and production sources as illustrated in Figure~\ref{fig:SmartGrid}.
In this context, EVs impose new burdens due to the extra demands they constitute, but also open opportunities thanks to the fact that their demands are relatively flexible, and that their batteries can be temporarily used to support the power grid: EVs can be active contributors in the smart grid instead of passive consumers. 

The important aspect stressed in this paper is that EVs cannot be assumed to be directly coordinated by a central entity controlling all charging processes. Indeed, EVs belong to individuals with specific preferences and constraints, who would not relinquish control of the charging process without being properly compensated. Instead, it is reasonable to assume that they react selfishly to management schemes: only when sufficient incentives are offered may EV owners coordinate their charging time and power, i.e., reschedule (directly or by giving some control to an external entity) the charging process rather than recharging their batteries within the shortest delay, which is convenient for them but problematic in the grid operator perspective. 
Those incentives can take several forms, from fixed rewards for letting the grid control the charging, to auctions for energy, or through time-varying prices set by grid operators.

Therefore, we think EV charging must be managed using market mechanisms, where participants are assumed to have different objectives. Hence an appropriate framework to study the EV management schemes is that of economy, and more precisely game theory~\cite{fudenberg1991game,Osborne1994course} which provides specific tools to model and analyze the interactions among self-interested actors.

This paper reviews the economy-driven schemes for EV charging management proposed in the literature. While the research on that topic is quite flourishing in the last years, there is to our knowledge no work presenting a comprehensive overview of the different approaches considered. This paper classifies the existing models, highlights their main assumptions and results, in order to compare them and identify the most promising types of mechanisms together with the directions that deserve further research.

EV charging management requires the support of a corresponding communication structure. In some algorithms, information is broadcasted from grid operators to EVs; bidirectional unicast is sometimes needed to coordinate the charging behaviors of specific EVs; finally EVs multicasting to charging stations (with or without station relaying) and stations responding (by unicast, multicast or broadcast) are necessary in reservation-based systems.

The importance of Information and Communication Technologies on the implementation of a so-called smart grid can never be overemphasized~\cite{Fan2013SGCommunications,Yan2013SGCommunications}, and specially designed communication systems for vehicles~\cite{Alsabaan2013VehicularCommun} are also relevant for better scheduling the charging of EVs. Hence charging algorithms and the corresponding communication systems should be considered simultaneously to make the best of their economical and environmental potentials. Existing works in the literature provide general overviews of the requirements and challenges; here we further investigate the economic properties of the charging algorithms, but keep track of their prerequisites on communication systems in terms of the volume and the frequency of information exchanges.

The remainder of this paper is organized as follows. 
Section~\ref{sec:EVsSystems} briefly discusses the technical environment of the charging problem, introduces the economic vocabulary and the desirable properties of an EV management scheme. 
The next two sections present and classify the charging schemes proposed in the literature to exploit the benefits and avoid undesirable outcomes from EVs entering the grid ecosystem: 
Section~\ref{sec:Unidirectional charging} focuses on \emph{unidirectional charging} (energy only goes from the grid to the EV batteries) while Section~\ref{sec:Bidirectional energy trading} allows \emph{bidirectional energy trading} (the grid can also take energy from the on-board EV batteries). 
Section~\ref{sec:Communication} summarizes the communication aspects of the schemes (type of exchanges, volume and frequency), while 
Section~\ref{sec:Mechanism} provides a general classification of all models and approaches, stressing their limitations to highlight the need for further research in specific directions. 
Section~\ref{sec:Conclusion} concludes the paper. 

%
%

\section{Techno-economic environment of EVs}
\label{sec:EVsSystems}

\subsection{Facilities for Electric Vehicle Charging}
\label{subsec:Facilities}
The term ``Electric Vehicle" can refer to a broad range of technologies. Generally speaking, the extension of this concept covers all vehicles using electric motor(s) for propulsion, including road and rail vehicles, surface and underwater vessels, even electric aircrafts. Since our paper concerns the charging management schemes and their impacts on the grid as well as on their owners from an economic perspective, we narrow the use of ``Electric Vehicle" to mention a passenger car with a battery that needs refills of electricity from external sources. 
Battery Electric Vehicles (BEV) and Plug-in Hybrid Electric Vehicles (PHEV) are two types of Plug-in Electric Vehicles (PEV); PHEVs differ from BEVs in that the former have a gasoline or diesel engine coexisting with an electric motor.

The economic mechanisms evoked in this paper mainly differ in the way prices are defined, in the mobility models (if any) of EVs, in the time scale considered, and in the directions for power flows (from the grid to EVs, or both ways). The specificities of EVs--being BEVs or PHEVs--do not play a major role with regard to the economic aspects, and often schemes are proposed that can be indifferently applicable to each type of EV. Hence in this survey we present mechanisms without always specifying the EV type; we do it when it has an influence on the performance or applicability of the scheme.

Note that charging can be performed in diverse ways: EVs can use an on-board or off-board charger~\cite{Yilmaz2012ChargingReview, Yilmaz2013Charging}, or use inductive charging while parked, thanks to Inductive Power Transmission (IPT) technology~\cite{Wu2011InductiveCharging, Khaligh2012Charging}. The ultimate experience of IPT is charging while in motion, of which a prototype named On-Line Electric Vehicle (\textit{OLEV}) has been designed in the Korea Advanced Institute of Science \& Technology~\cite{Shin2012OLEV}. Those cases being rare, we can consider in this paper that the charging is done via a physical connection with an on-board plug. 

To insure safe electricity delivery to an EV from the source, some particular EV Supply Equipment (EVSE) is needed, which puts tight constraints on how EVs can be recharged (or discharged if possible). 
The charging rate limit, battery capacity and AC/DC conversion efficiency vary among the different charging facilities and patterns. 
Two levels for AC charging and three levels for DC charging are approved by the SAE J1772 standard\footnote{http://www.sae.org/smartgrid/chargingspeeds.pdf}, as shown in Table~\ref{fig:SAE_J1772}, giving the estimated time $T$ needed to fully recharge a battery with $25$kWh usable pack size, starting from an initial State Of Charge (SOC) of $20\%$.
\begin{table}[htbp]
   \caption{Charging powers and corresponding charging duration $T$ under the SAE J1772 standard}
   \label{fig:SAE_J1772}
\begin{tabular}{c|c|c|}
\cline{2-3}
&Level 1&Level 2\\
\cline{2-3}
~~~~~AC& $\sim1.9$kW & $\sim19.2$kW\\
\cline{2-3}
&$T=17$h&$T=1.2$h\\
\cline{2-3}
\end{tabular}
\vspace{2ex}
\\
\begin{tabular}{c|p{.17\textwidth}|c|c|}
\cline{2-4}
&\centering Level 1&Level 2&Level 3\\
\cline{2-4}
~~~~~DC&\centering $\sim36$kW & $\sim90$kW& $\sim240$kW\\
\cline{2-4}
&\centering$T<1$h&$T<20$min&$T<10$min\\
\cline{2-4}
\end{tabular}
\end{table}
There are other charging standard proposals, which roughly correspond to the categories in Table~\ref{fig:SAE_J1772}. For example CHAdeMO\footnote{\url{http://www.chademo.com}} falls into DC Level 2, and Tesla Supercharger overlaps with DC Level 3. 

\begin{figure}[htbp]
   \centering
   \includegraphics[width=.5\textwidth]{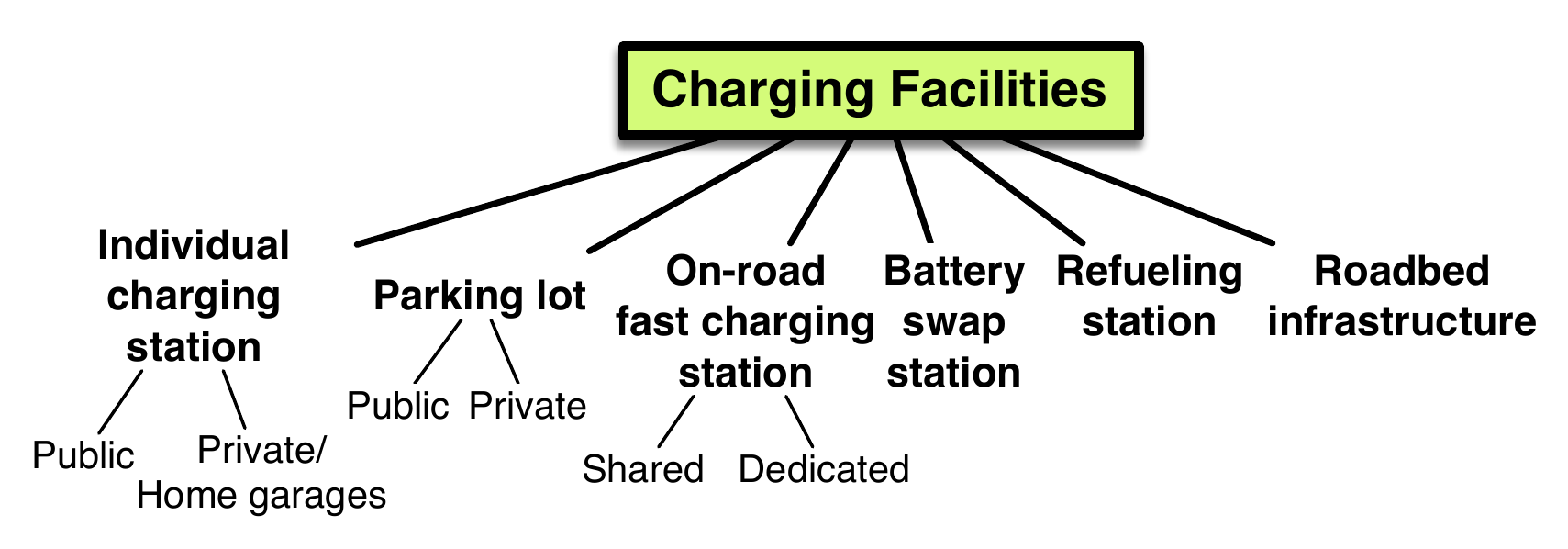} 
   \centering
   \caption{Classification of the charging facilities for EVs}
   \label{fig:charging_facilities}
\end{figure}

At the other end of the wire stretching out from an EV socket is a charging station. 
Figure~\ref{fig:charging_facilities} summarizes the main categories in which we can divide the charging stations. 
\textit{Individual stations} capable of charging a single EV refer to those located in individual homes. \textit{Parking lots} for EVs are yet to be developed to their full potential: they contain many individual EVSEs in physical proximity, belonging to the same entity. Public EV parkings are open to any EV, while private EV parkings provide access only to a specific fleet of EVs, e.g., owned by a single company. 
\textit{On-road stations} are relays for EVs on long journeys, they can generally charge EVs at the highest possible rate to minimize the delay.

\textit{Roadbed infrastructures} for EVs are based on IPT technology~\cite{Lukic2013Inductive}. We already witness roadbed infrastructures that charge EVs at traffic intersections~\cite{Mohrehkesh2011Wireless} or even without stopping~\cite{Lee2010OLEV}. As some EVs can use other types of energy sources, they can be replenished in \textit{refueling station}, e.g. classical petrol stations, compressed air stations, or \textit{battery-swapping stations}. 
Those charging solutions are out of the scope of this paper due to the fact that they are either to some extent overlapping with refueling problems for conventional cars, or still in experimental stage.


\subsection{Electric Vehicles -- An enabler of the Smart Grid and a participant in Electricity markets}
\label{subsec:Market}

The Smart Grid is an evolution of the Power Grid which is expected to lead to a more efficient use of the grid resources, for example with a reduced Peak-to-Average power consumption ratio, faster repairs, self-healing and self-optimizing possibilities, and full integration of renewable energy sources.

Demand Response (DR) is the possibility for the power grid to alter the consumption patterns of end users; it can be implemented through various mechanisms. DR was initially used primarily toward large electricity consumers, but the transition to the Smart Grid provides a paradigm shift, where every load, no matter how small, can participate in a DR program. 
Energy Storage is a key technology for the integration of Renewable Energy Sources to the grid. Pumped-storage hydroelectricity (PSH) accounts for 99\% of the world bulk storage capacity\footnote{\url{http://www.economist.com/node/21548495}}, but there are physical limitations to the quantity of energy that these types of storage can hold. 

Electric Vehicles can both participate in DR and serve as Energy Storage facilities. They can respond to DR signals, such as price variations or direct control messages by modulating their power consumption, thus providing necessary flexibility to the grid operator. In some cases, EVs can also inject electricity back to the grid, thus serving as distributed energy sources. These can be leveraged by the network operators to improve renewable energy integration, to help self-healing or to provide ancillary services, so as to reduce the dependency on specialized equipments like diesel generators.

\begin{figure}[htbp]
   \centering
   \includegraphics[width=9cm]{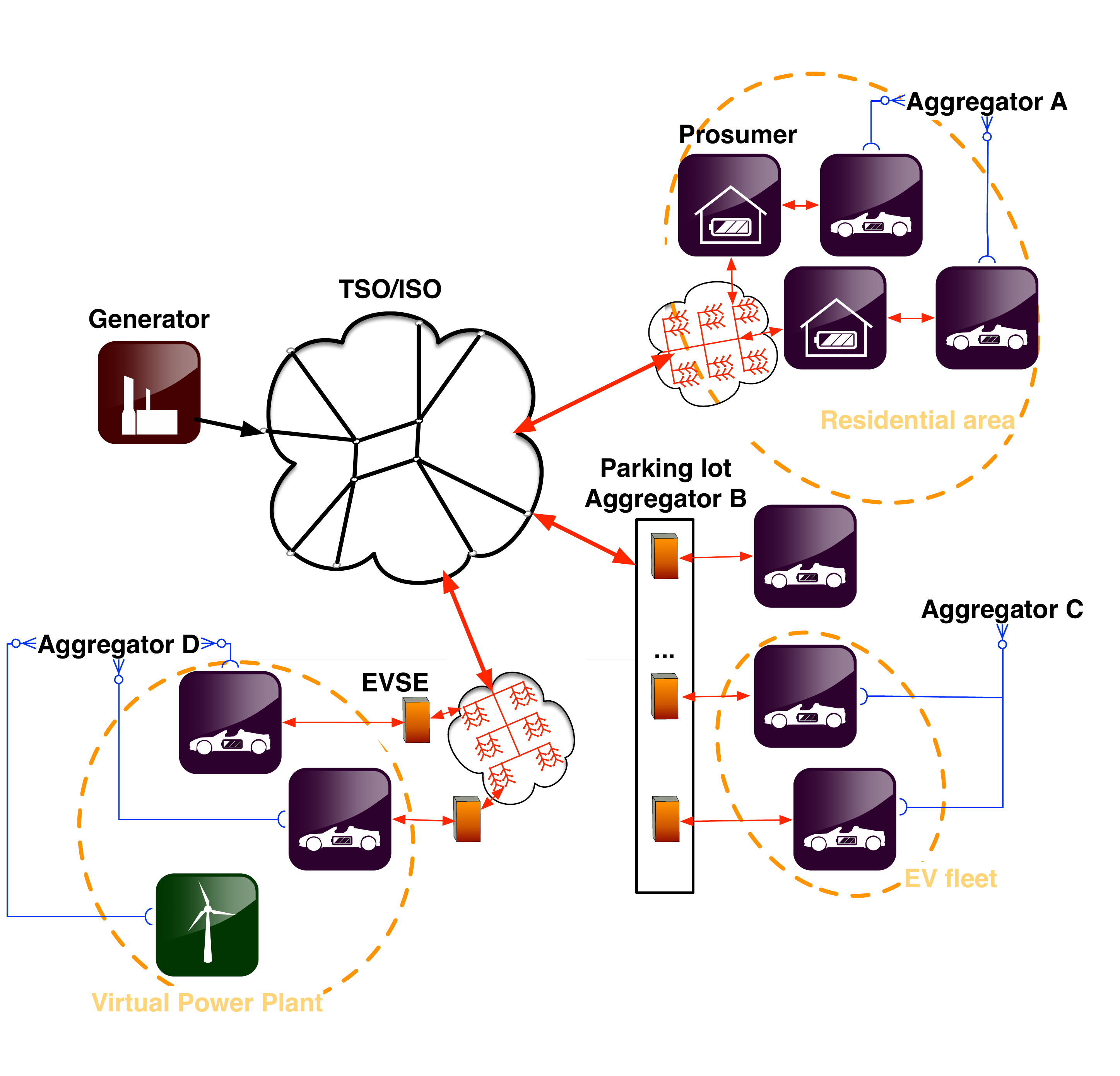} 
   \centering
   \caption{Smart Grid Actors related to EV charging.}
   \label{fig:EVactors}
\end{figure}

Figure~\ref{fig:EVactors} shows the major entities related to EV charging. 
A Transmission System Operator~\cite{DefTSO} (TSO, in Europe)--or in some contexts (in North America) an Independent System Operator~\cite{DefISO} (ISO)--is responsible for operating, ensuring the maintenance of and, if necessary, developing the transmission system in a given area.
Consumers equipped with energy sources that can deliver electricity to the distribution network are called prosumers.

In a classical electricity market, end-users have contracts with an electricity \textit{retailer}, who buys the electricity produced by \textit{generators}. 
The transaction can be brokered via a bilateral agreement or on a wholesale market. 
As the aggregated energy consumption of a big region can be known with satisfactory precision well in advance, contracts for buying the bulk of the necessary electricity can be done a year or a month ahead on the \textit{futures market}. 
However, electricity consumption is heavily dependent on the weather, thus requires a significant amount of energy to be traded $24$ hours in advance on the \textit{day-ahead market}. 
Finally, fine adjustments can be made up to an hour ahead, which are traded on the \textit{intra-day market}. 

To match supply and demand for electricity instantaneously, ISO/TSOs operate ancillary services markets (generally using auctions) where they purchase ancillary services from generators and/or consumers who have the ability to vary their generation or consumption powers. ISO/TSOs also keep a close watch on the efficiency and effectiveness of those markets.

\subsection{Dealing with self-interested actors}
\label{subsec:Self-interested}
As elaborated before, EV charging involves many Smart Grid actors, whose objectives are not necessarily aligned: EV owners want to store enough energy as quickly as possible, and at the lowest cost, whatever the impact on other EVs or on generation costs; electricity producers and retailers are mainly driven by net benefits; while ISOs generally aim to ensure the most efficient use of resources and to maintain the supply-demand balance.

Therefore, when designing mechanisms to decide allocations and prices paid, one has to anticipate that the actors may try to play the system at their advantage. For example, if decisions are made based on signals from users such as their willingness-to-pay, the rules should ensure that reporting untruthful values does not bring any gain to the corresponding actors: such a property is called \emph{incentive compatibility}.

More generally, an appropriate framework to study the interactions among several decision-makers is that of game theory~\cite{fudenberg1991game}. A key notion is the Nash equilibrium, that is an outcome (a decision made by each actor) such that no actor can improve his individual payoff (utility) through an unilateral move. As stable situations, Nash equilibria are often considered to be the expected outcomes from interactions. Hence many of the mechanisms described in this paper rely on that notion.

Nash equilibria can be attained when all actors have perfect knowledge of their opponents, their decision sets, and their preferences. But those strong (and often unrealistic) conditions are not necessary: in several cases the Nash equilibria can be reached or approached via some limited information exchanges among actors, or even without such exchanges but just by trying out decisions and \emph{learning} the best ones~\cite{fudenberg1998theory}.

To summarize the EV charging problem setting, we recall the relevant actors and set up the vocabulary as below:

\begin{itemize}

\item EV: A physical electric vehicle or its owner who will generally be assumed to have a \emph{utility function} (or benefit), that represents his preferences. We will mostly use the classical \emph{quasi-linear utility} model~\cite{mankiw2014principles}: for a given price and energy allocation, the EV owner utility will be the difference between the owner's \emph{willingness-to-pay} (or \emph{valuation}, i.e., the value of energy for him, expressed in monetary units) and the price actually paid.

\item Aggregator: An entity acting as an intermediary between the demand (retailers/users) and supply (generators, ISO/TSO or charging stations in some scenarios) sides of the electricity market \cite{Gkatzikis2013role}. When an aggregator is designed to be a representative of a group of EV owners, its utility will be the aggregated user utility. Otherwise, when it acts in its own interest as an intermediate energy supplier, the measure of utility will similarly be the difference between \emph{revenues} (the monetary gains from their clients) and \emph{costs}. That difference is often called \emph{benefit}.

\item EV charging station: The owner and/or operator of one or several EVSEs in physical proximity, who allows EV recharging and/or discharging with the aim of maximizing revenue, but always under some physical constraints such as local transformer capacity and standard recharging power level. 

\item ISO (or TSO): An entity in charge of operating and maintaining the transmission system in a given area. It sets a constraint for the aggregated EV load according to the transformer capacity, and purchases ancillary services when necessary, in order to maintain the supply-demand balance.

\end{itemize}

The aggregated utility of all users (here, EVs) is called \emph{user welfare}, and the aggregated utility of all suppliers (EV charging stations or aggregator) is the \emph{supplier welfare}. \emph{Social welfare} (=user welfare+supplier welfare) quantifies the global value of the system for the society, and is computed as the sum of all users' valuations minus all costs (production, transportation, if any). Note that money exchanges do not appear in that measure, since they stay within the society.

To provide a guideline for future proposals, we list in Table~\ref{tab:criteria} the main questions raised by EV charging, and summarize from our point of view, the criteria that make a good charging management scheme. Also, we indicate in which sections of this paper those points are addressed.

\begin{table*}[htbp]
   \centering
   \topcaption{Main questions related to the EV charging problem, and desirable properties}
   \label{tab:criteria}
   \begin{tabular}{|p{6.5cm}|p{6.5cm}|p{1.3cm}|p{2cm}|}
   \hline
   Question & Criteria & Examples & In this paper \\
   \hline
   \hline
    How to settle the conflict between EV demand and grid capacity? & A desirable management scheme achieves higher EV satisfaction and/or station revenue, meanwhile lowers the grid burden.&\cite{Galus2008DSM,Galus2009EnergyNetworks,Samadi2010DSM,Tushar2012Economics,Maille2006Auctions,Ardakanian2014DistributedControl,Qin2011ChargingWaiting,Guo2014Rapid,Liu2014Routing,Hu2014Congestion,Hu2015CongestionNetwork,Franco2015MILP} & Section~\ref{subsubsec:SharingUsers}, \ref{subsubsec:SharingFuture}\\
   \hline
   How to coordinate the time-flexible EV demand scattered in individual EVs to perform load shedding, peak shaving, and to smooth renewable energy output? & A well designed pricing policy can incentivize participants to shift their demands in a distributed manner, without intrusively taking full control over their charging processes.& \cite{Beaude2012ChargingGame,Rad2010DSM,Ma2013Valley,Gan2013OptimalProtocol,Sortomme2011ChargingStrategies,Bessa2012Bidding,Leterme2014Wind,Vasirani2013VPP,Xie2012Wind,Binetti2015SRTG} & Section~\ref{subsubsec:SharingSlots}, \ref{subsubsec:UniRegulation}, \ref{subsec:V2GStorage}\\
   \hline 
   How could an EV owner reduce his/her electricity expenses by paying the time-of-use electricity price? & A satisfying charging program is flexible in order to respect EV owners' travel plans, and is robust to price uncertainty.& \cite{Conejo2010DRM,Hutson2008Intelligent,Liang2012V2GStorage,Han2010OptimalV2G,Liang2013EnergyDelivery,Yang2014Taxi} & Section~\ref{subsec:Arbitrage}\\
   \hline
   How to dispatch ancillary service tasks as well as the associated revenue among EVs providing such services? & A good allocation satisfies some fairness properties in terms of actor utilities. & \cite{Quinn2010V2G,Kamboj2010V2G,Kamboj2011VandE,Garzas2012FairDesign,Sun2013RegulationAllocation,Sun2014RegulationAllocation,Wu2012V2G,Wu2012Wind,Gao2013ContractRegulation} & Section~\ref{subsec:V2GRegulation}\\
   \hline
   How to organize the EV charging market between self-interested EV owners and revenue-pursuing charging stations? & A good mechanism should  be incentive compatible and achieve high (near-optimal) user/supplier/social welfare. & \cite{Gerding2011MechanismDesign,Gerding2013Mechanism,Qin2011ChargingWaiting,Guo2014Rapid,Liu2014Routing,Garzas2012StationSelection} & Section~\ref{subsec:Dynamic} \\
   \hline
   \end{tabular}
\end{table*}

%
%

\section{Unidirectional charging mechanisms}
\label{sec:Unidirectional charging}

In this section, we assume that energy can only go from the grid to EV batteries.
Electricity is expensive to store and supply over the grid must match demand at all instants: hence it is not possible for the grid to simply produce in anticipation the power needed to satisfy the charging requests that will occur from possibly many EVs over some periods of time. 
Standby generation units can be swathed on, but incur high costs, hence this is not a satisfying solution either. 
Remark that the generation part is not the only limiting factor: transmission networks and transformer station limits constitute other bottlenecks.
We therefore consider here the scenario where several EVs are plugged-in for recharging, but the available energy is not sufficient to feed them all (or producing extra energy incurs high costs), so the aggregator is responsible for allocating the scarce resource among the clients. 

This section reviews the main economic approaches to manage the (unidirectional) charging of EVs. 
We first describe \emph{static} approaches for energy sharing (where the objective and decisions are based on a snapshot of the system regardless of possible impacts of future variations),
then extend the sharing problem to \emph{dynamic} scenarios (where the uncertainty of future events is taken into account); we also consider the mobility aspects of EVs (involving the choice of locations to charge, and price/distance tradeoffs) and finally point out mechanisms based on \emph{frequency regulation}.

\subsection {Static unidirectional recharging}

\subsubsection{Sharing energy efficiently among users}
\label{subsubsec:SharingUsers}
%
This subsection is devoted to the energy allocation problem within one indivisible time slot, i.e., only the current demands are considered and there is no uncertainty considered about future events (variations in supply and/or demand). We start with topology-free models, where each EV's consumption is constrained by its charger and battery, then together with other EVs, jointly curbed by the supply (typically, from their common aggregator). Then we move on to topology based models, where the throughput of the transformers further narrows the feasible choices.

\paragraph{Sharing without topology-based constraints}
\label{para:WithoutTopology}
Consider a charging station with several plugged-in EVs demanding electricity. How should the station dispatch the scarce available energy among them? We suppose here that there is no discrimination among EVs caused by the topology of the (sub-)grid they are connected to. Energy supply is considered as a  constant for models in~\cite{Galus2008DSM,Galus2009EnergyNetworks}, and as a variable in~\cite{Samadi2010DSM,Tushar2012Economics}.

%
Galus and Andersson~\cite{Galus2008DSM} consider a large amount of PHEVs connected to an energy hub which converts gas and electricity to cover a commercial area's heat and electricity needs. 
Hence the total energy available to PHEVs is the transferring limit of the hub, minus the commercial area's un-shiftable demand. Each PHEV is assumed to report truthfully to the aggregator an individual (utility) parameter describing its willingness-to-pay for one unit of energy, at every time instant. 
This value depends on the gap between the current SOC and its target, as well as the time left before its departure. 
The aggregator then dispatches the available power, based on those parameters collected from all plugged EVs, to maximize the total (declared) value of energy for PHEVs, generally feeding first the EVs with lower SOC and imminent departure. A strong assumption made here is that EV owners do not try to play the system by falsely declaring their utility parameters to obtain higher utilities. The authors extend their work by adding a network operator, in charge of a higher-level dispatch of electricity and gas over all the aggregators~\cite{Galus2009EnergyNetworks}, thus the supply limit is simultaneously restricted to the capability of the hub and the electricity and gas fed-in to an aggregator by that network operator. 

%
In contrast to~\cite{Galus2008DSM} where energy supply is given as a constraint, Samadi~\textit{et al.}~\cite{Samadi2010DSM} let the aggregator decide the amount of electricity to sell in order to maximize social welfare, that is the aggregated benefit of all the self-interested users minus the generation cost. 
They propose a distributed iterative algorithm where the aggregator updates the unit energy price and each user responds by updating his load (to the utility-maximizing one under the present price) until convergence, at which point 
energy allocations become effective. 
Here again, no strategic behavior from users is assumed: they react myopically without integrating the fact that their utilities depend only on the converged outcome.

%
In Tushar~\textit{et al.}'s model~\cite{Tushar2012Economics}, users are not only informed of the price, but also of the total consumption limit. 
Each user aims to maximize his utility function, while knowing that if total demand exceeds the consumption limit, then none will be allocated any electricity. 
This scenario is modeled as a Stackelberg game~\cite{Osborne1994course} (also known as leader-follower game), with the aggregator as the leader, setting prices so as to maximize revenue; and EVs as the followers--price-takers competing for resource through their demands. 
The leader sets the price first, then the followers send their demand to an intermediary manager, until the unique EV equilibrium for that price is reached. 
The total consumption is then sent to the leader, who updates the price to achieve a higher revenue; that process being repeated until the revenue is maximized. 
%

%
\paragraph{Sharing with topology-based constraints}
\label{para:WithTopology}
The following models share the assumption that EVs are connected at the leaves of a tree-like distributed network. The objective of an allocation can be efficiency~\cite{Maille2006Auctions} or fairness~\cite{Ardakanian2014DistributedControl}.

Maill\'e and Tuffin~\cite{Maille2006Auctions} propose a solution to share resource among self-interested users over a tree structure, through an auction and with the objective of maximizing social welfare. 
The mechanism was initially defined for bandwidth sharing in telecommunication access networks, but is also applicable to energy: an EV can send several bids to the auctioneer, each with the form of a $(\text{unit\_price,quantity})$ pair; the auctioneer then computes energy allocations and prices based on the bids submitted by all  EVs. The number of pairs one EV can submit is chosen as a trade-off between efficiency and (communication and computational) complexity. 
The mechanism in~\cite{Maille2006Auctions} follows the principle of Vickrey-Clarke-Groves mechanisms~\cite{vickrey1961counterspeculation,clarke1971multipart,groves1973incentives}; it incentivizes truthful bidding for the users and guarantees efficient allocation--in the sense of user welfare maximizing, since no costs are assumed here. 

%
Rosenberg and Keshav~\cite{Ardakanian2014DistributedControl} aim at finding a proportionally fair~\cite{KELLYF1997fair} sharing of a fixed amount of energy among users. 
The algorithm consists in each link computing its congestion or \emph{shadow price}~\cite{kelly1998rate}, and transmitting downwards the total congestion price from the root of the tree (wherefrom energy is available) to users plugged at leaves; the latter then demand their utility-maximizing amount after receiving the price (assuming logarithmic utility functions). 
Such a method converges to the proportionally fair allocations. 
Note that users here are not aware of the links capacity limits, so their initial demands might exceed them before reaching convergence, an outcome not occurring in~\cite{Tushar2012Economics} where users sharing a link know its capacity and act to avoid outstripping it.

\paragraph{Example}
\label{para:Example}
Now we illustrate some of those approaches via a simple example. 

Consider an aggregator having to allocate energy to two users $A$ and $B$ with (non-decreasing) concave quadratic valuation functions $\theta$ (indicating their willingness-to-pay for energy) as expressed below:
\begin{equation}\label{benefit function}
\theta(x)=
\begin{cases}
- ax^2 + bx & x\le \frac{b}{2a}\\
\frac{b^2}{4a} & x>\frac{b}{2a},
\end{cases}
\end{equation}
%
where $x$ is the allocated energy and $a,b$ are user-specific parameters. 
Note that $\frac{b}{2a}$ is the maximum amount of energy that the user wants, i.e., giving him more than this value won't increase his valuation. Users $A$ and $B$ differ in their preferences: set $a_A=0.5, a_B=1$ (the respective values of parameter $a$ for player $A$ and $B$) and $b_A=b_B=2$. 
The utility of each player is therefore the difference between his valuation function, and the price he is charged (typically, $px$ with $p$ denoting the unit price).

%
The aggregator acts as a representative of the EVs in~\cite{Galus2008DSM,Maille2006Auctions}, trying to maximize the aggregated user utility. Similarly, in~\cite{Ardakanian2014DistributedControl} the aggregator has also a user-based objective, namely proportional fairness.
In contrast, in~\cite{Tushar2012Economics,Samadi2010DSM} he plays ``against'' EV users, trying to maximize his revenue by setting the unit price $p$. The supply constraint is a tight bound of $C$ in~\cite{Tushar2012Economics}, while in~\cite{Samadi2010DSM} it is part of the decision variables, the authors assume a cost of $\alpha C^2$ and consider $C$ to be optimized by the aggregator.

%
Table~\ref{tab:toy_example} shows the outcomes of those approaches for our example.
Remark that welfare-oriented approaches \cite{Galus2008DSM} (and~\cite{Maille2006Auctions} if $C_2\geq C_1$ for example) lead to the same allocations as the revenue-oriented ones~\cite{Tushar2012Economics,Samadi2010DSM}. 
Those allocations correspond to demands at the \emph{market price} (the unit price as which demand equals supply); indeed such allocations are efficient, but also allow the aggregator to extract the maximum surplus from users. 
Note however that the prices paid are different: with VCG-based schemes, users are charged below the market price, which can be interpreted as the cost for having them reveal truthfully their valuation (while this information-revealing aspect is not considered in~\cite{Tushar2012Economics,Samadi2010DSM}). 

\begin{figure}[htbp]
   \centering
   \includegraphics[width=8cm]{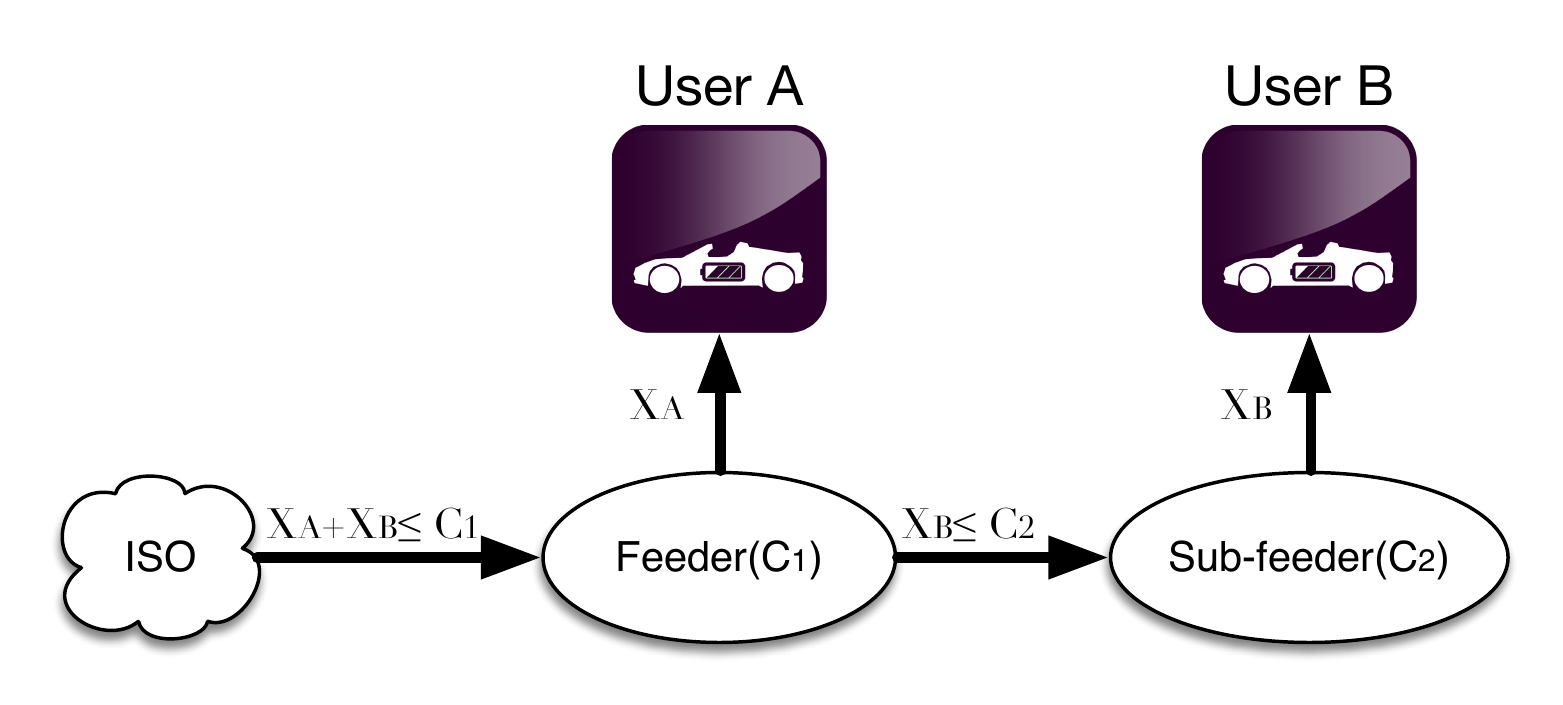} 
   \caption{Capacity constraints: a simple tree topology.}
   \label{fig:Alice&Bob}
\end{figure}

%
For the models in~\cite{Ardakanian2014DistributedControl,Maille2006Auctions}, that consider tree-like network topologies, we take in our example the simple topology of Figure~\ref{fig:Alice&Bob}, where $C_1$ and $C_2$ are capacity limits. 

%
The objective in~\cite{Ardakanian2014DistributedControl} is to achieve proportional fairness in this setting, or equivalently, to maximize $\log{x_A} + \log{x_B}$ under the capacity constraints. Hence energy is shared equally if the constraints allow it, as shown in Table~\ref{tab:toy_example}. 

\begin{table*}[htbp]
\centering
\caption{Comparing mechanisms proposed for static scenarios on a toy example} \label{tab:toy_example}
\begin{tabular}{|c|c|c|c|c|}
\hline
 & Aggregator objective & Constraints & Allocation $x_A$ & Allocation $x_B$ \\
\hline
\hline
\cite{Galus2008DSM} & $\displaystyle\max_{x_A,x_B} \theta_A(x_A)+\theta_B(x_B)$ & $x_A + x_B \le C$ & $\min(1,2C/3)$ & $\min(1/2,C/3)$  \\
\hline
\cite{Tushar2012Economics} & $\displaystyle\max_p p(x_A+x_B)$ & $x_A + x_B \le C$ & $\min(1,2C/3)$ & $\min(1/2,C/3)$  \\
\hline
\cite{Samadi2010DSM} & $\displaystyle\max_{p,C} p (x_A+x_B) - \alpha C^2$ & $x_A + x_B \le C$ & $\frac{2}{3}C=\frac{2}{2+3\alpha}$ & $\frac{1}{3}C=\frac{1}{2+3\alpha}$ \\
\hline
\cite{Maille2006Auctions} & $\displaystyle\max_{x_A,x_B} \theta_A(x_A)+\theta_B(x_B)$ & $x_A + x_B \le C_1, x_B\leq C
_2$ & $\min\left(1,\max(\frac{2C_1}{3},C_1-\min(\frac 12,C_2))\right)$ & $\min(\frac 12,\frac{C_1}{3},C_2)$  \\
\hline
\cite{Ardakanian2014DistributedControl} & $\displaystyle\max_{x_A,x_B} \log{x_A} + \log{x_B}$ & $x_A + x_B \le C_1, x_B \le C_2 $ & 
$C_1-\min(C_1/2,C_2)$& 
$\min(C_1/2,C_2)$\\
\hline
\end{tabular}
\end{table*}



\subsubsection{Electricity sharing over several time slots}
\label{subsubsec:SharingSlots}
This subsection adds the dimension of time when scheduling EV charging. A given time interval is divided into multiple time slots. Unlike the previous subsection which treats as decision variables the amount of electricity to be allocated among EVs, in this subsection those variables now expand on time, becoming vectors, to exploit the time flexibility of allocation. Therefore the problem is to reshape the aggregated charging \emph{load curve} under constraints on the total energy transferred. We start with models aiming at forming a flat load curve, then turn to those that can shape the load into an arbitrary curve.

\paragraph{Flat charging curves}
\label{para:Flat}
Time is discretized so a charging plan for an EV is a vector over slots, with the magnitudes representing the charging rates. This rate takes discrete values in~\cite{Beaude2012ChargingGame,Binetti2015SRTG} and continuous values in~\cite{Rad2010DSM}.

%
Beaude, Lasaulce and Hennebel~\cite{Beaude2012ChargingGame} slice one time period (typically one day) into several slots (e.g., of length 30 minutes), and users choose when to start recharging their EVs at a constant power level without interruption until reaching their target SOCs. 
In other words, the charging demand is a shiftable rectangle covering several slots. 
For the aggregator, a supply increase causes a cost increase, and the cost function is assumed to be continuously differentiable and strictly convex. This cost is directly transferred through prices to users, who are aware of this mechanism and want to choose the best charge starting slot(s) to minimize their individual costs. 

%
Let us illustrate the scheme through a simple example.
Suppose EV $A$ (resp., $B$) needs a one-time-slot recharging at the power of $c$. They can choose between slot $1$ and slot $2$. 
Denote the consumption profile for $A$ (resp., $B$) in time slots $1$ and $2$ with the vector $[x_A^1\quad x_A^2]$ (resp., $[x_B^1\quad x_B^2]$). 
For a specific time slot, the aggregated load can be $2c$, $c$, or zero depending on user decisions, with respective costs to the aggregator $\text{Cost}(2c) > \text{Cost} (c) > \text{Cost} (0)$. This cost is handed down to the users in the form of unit prices $p(\text{Cost}(2c)) > p(\text{Cost}(c)) > p(\text{Cost}(0))$.

%
The game played among users is proved to be a potential game~\cite{monderer1996potential}, hence having pure Nash equilibria. Not all those equilibria yield identical cost, but the authors prove that for an infinite number of cars the equilibrium is unique and optimal. 

%
In the same vein, Mohsenian-Rad~\textit{et al.}~\cite{Rad2010DSM} use this consumption-dependent electricity price to elicit users to voluntarily minimize the cost to the aggregator, and meanwhile reduce the peak-to-average ratio of the load curve. 
The aggregator sets a unit price linearly increasing in the consumption level, so that the price paid is quadratic in the consumption. 
Users have multiple independent appliances to manage, and the constraint of non-stopping recharging in~\cite{Beaude2012ChargingGame} is relaxed, so higher flexility is offered: the charging rate is variable as long as the total energy injected to one appliance meets the client's demand. 
When a day starts, each user first starts from a random hourly consumption schedule and broadcasts it to the rest of the community. Then, sequentially, users choose their cost-minimizing schedules based on those received from the others and their own daily needs. The authors prove that the process converges, to a unique equilibrium where the total charging cost is minimized. A desirable byproduct of minimizing the cost is that the peak-to-average ratio of the load curve is also significantly reduced: although the solutions of these two problems are not identical, data analyses suggest that they are close, since the lowest achievable peak-to-average is only 0.05\% lower than that achieved by the cost-minimizing solution. 

The convergence requires rounds of bi-directional communication between an EV and the aggregator. To accelerate the procedure and achieve real-time responds, Binetti~\textit{et al.}~\cite{Binetti2015SRTG} propose to schedule one EV at a time, once it connects to the grid. First the aggregator anticipates its load curve for the following 24 hours, and lets the first arriving EV know about this profile upon arrival; then the EV owner, after a simple computation, decides when to start recharging its EV at a constant yet self-defined power level, without interruption. The computation complexity is low and can be easily adapted to the circumstance where EVs arrive in a batch. In~\cite{Binetti2015SRTG} EV owners arrange their recharging with the aim of minimizing the objective function of the aggregator, which is a linear combination of the variance and peak of the aggregated load profile, thus users are assumed altruistic; but a more realistic approach should cover EV owner selfishness, hence an incentive problem: how to define prices so that selfish users behave in the best interest of the aggregator? We expect load-dependent prices to lead to situations where the aggregator cost is (at least approximately) minimized, as is done under other assumptions in~\cite{Beaude2012ChargingGame,Rad2010DSM}.

\paragraph{Following an arbitrary curve}
\label{para:Arbitrary}
Merely flattening the aggregated charging consumption of EVs is not always desirable or sufficient, especially when EVs share the supply system with other consumers. To flatten the overall demand curve, the EV consumption should be adjusted according to the external un-shiftable loads. Following are examples of guiding EVs through the electricity price, so that their aggregated load follows a predefined curve. 

%
Ma, Callaway and Hiskens~\cite{Ma2013Valley} model the behavior of self-interested users as a noncooperative game, the objective of the charging control being \emph{valley filling}, i.e., shifting EV demand to the valley hours of the non-EV load. 
A consumption-dependent electricity price (a linear function of the ratio of the real-time consumption to the generation capacity) elicits users to defer their charging process toward the valley periods, where prices hit the bottom. 
To avoid oscillations in user behavior and ensure convergence, 
an extra fee is added to the electricity price as a penalty on deviation from the population average, so that users selfishly minimizing their costs converge to a Nash equilibrium, which happens to be the socially optimum outcome if all EVs have an identical charging deadline. By adopting a different form of penalty, Gan, Topcu and Low~\cite{Gan2013OptimalProtocol} prove the convergence to the optimum for EVs with different deadlines. Moreover, they extend the algorithm so that the aggregate load can follow any given profile, hence its application goes beyond valley filling. 

When real-time electricity prices can reflect the congestion status, an EV would be contributing to valley-filling by simply following a cost-minimizing charging program. Franco, Rider and Romero~\cite{Franco2015MILP} seek a daily charging dispatch that achieves cost minimization under hourly electricity prices. They consider a specific distribution network where each node brings a constraint about the consumption it can support. The aggregator solves the problem in a centralized manner, i.e. tries to postpone the shiftable EV loads to the time slots with lower prices, while respecting the constraints and satisfying EV demand. 
Similarly, Hu~\textit{et al.}~\cite{Hu2014Congestion} propose a centralized cost-minimizing control mechanism based on predicted hourly electricity prices, where the aggregator directly controls the charging of each plugged EV, whose daily travel plan and corresponding energy demand can be estimated day-ahead. The aggregators sharing a distribution grid respond to hourly congestion prices set by the ISO, by updating their previously optimized EV recharging schedules. After convergence, the ISO re-sets the price depending on its supply capacity, until the overall energy consumption scheduled of all the aggregators falls below this capacity. The authors recently extended this work in~\cite{Hu2015CongestionNetwork}, to the case of a tree-like distribution network where EVs are plugged on the leaves.

\subsubsection{Summary of Static unidirectional recharging}
%
Table~\ref{tab:Static unidirectional recharging} summarizes the static approaches, differentiating them according to the type of economic model considered, the controller's objective, and the main model constraints. The first group (\cite{Galus2008DSM,Samadi2010DSM,Tushar2012Economics}) uses Stackelberg game models, with the aggregator being the leader and EVs the followers. The leader plays with the electricity price and followers adapt their consumption. This method can be used to achieve different objectives, such as user welfare and social welfare, in an iterative manner. When topology-based constraints are considered (\cite{Maille2006Auctions,Ardakanian2014DistributedControl}), EVs might not be aware of the whole topology and/or constraints of each segment, but congestion information at each node is handed down in the form of electricity prices. This method can achieve proportional fairness among homogeneous cooperative users~\cite{Ardakanian2014DistributedControl}. For heterogenous self-interested users, each with a private utility function, the central controller can organize an auction and dispatch energy efficiently among bidders, respecting topology constraints~\cite{Maille2006Auctions}.

All those charging schemes consider imposing consumption-dependent electricity prices to cost-sensitive users. 
While user demands are assumed elastic in schemes studying a single time slot, they are considered fixed for those designed for several time slots, the flexibility stemming from the repartition of consumption over time to meet demand constraints. That fixed-demand assumption is mathematically convenient (in particular, the optimal load curve is unique and computable), but it ignores the fact that EVs may benefit from alternative energy sources and therefore have flexible demand for grid power.
So we encourage future research to consider 
demand flexibility in both time and volume. This complicates the analysis of the aggregator's task (to choose a load curve) and of the EV choices (among the different sources), but we believe it is worth studying.

One inherent difficulty in distributed systems is convergence. Although it is mathematically convenient to assume arbitrarily variable charging rates between slots, batteries actually prefer stable charging rates. 
This hinders the convergence to global optimum in atomic charging games, and results in optimality being only achievable for an infinite number of EVs~\cite{Beaude2012ChargingGame}. 
Convergence can be guaranteed by modifying user choices (e.g., through penalties as in~\cite{Gan2013OptimalProtocol}), but this must translate into economic incentives by affecting utilities, hence comes with a cost.
\begin{table*}[htbp]
\centering
\topcaption{Economic approaches for static unidirectional recharging}
\begin{tabular}{|c|c|c|l|}
\hline
Paper & Model &  Aggregator objective &Constraints \\
\hline
\hline
\cite{Galus2008DSM}& Stackelberg game & User welfare& \begin{tabular}{l}Fixed produced energy\end{tabular}\\
\hline
\cite{Tushar2012Economics} & Stackelberg game & Revenue  & \begin{tabular}{l}Fixed produced energy\end{tabular}\\
\hline
\cite{Samadi2010DSM} & Stackelberg game & Revenue minus costs & \begin{tabular}{l}Quadratic cost for produced energy\end{tabular}\\
\hline
\cite{Ardakanian2014DistributedControl} & Cooperation & Proportional fairness  & \begin{tabular}{l}Fixed produced energy\\Fixed transmission capacities\end{tabular}\\
\hline
\cite{Maille2006Auctions} & Auction &  User welfare &  \begin{tabular}{l}Fixed produced energy\\Fixed transmission capacities\end{tabular} \\
\hline
\cite{Beaude2012ChargingGame} & Potential game & Generation and delivery costs & \begin{tabular}{l}Fixed charging rate per EV\\No interruption of charging\end{tabular}\\
\hline
\cite{Rad2010DSM}& Potential game & Energy cost (and Peak-to-Average ratio) & \begin{tabular}{l}Fixed min and max consumption for appliances and EVs\end{tabular}\\
\hline
\cite{Ma2013Valley}& Non-cooperative game 
& Energy cost (also valley filling)&\begin{tabular}{l}Fixed non-EV demand\end{tabular}\\
\hline
\cite{Gan2013OptimalProtocol} & Cooperation & Vally filling & \begin{tabular}{l}Fixed non-EV demand\end{tabular}\\
\hline
\end{tabular} 
\label{tab:Static unidirectional recharging}
\end{table*}



\subsection{Dynamic models}
\label{subsec:Dynamic}
\subsubsection{Dealing with uncertainties about future events}
\label{subsubsec:SharingFuture}

%
All the models described so far are \emph{static}, in the sense that they consider a time interval (be it one time slot or several) where all the information needed to find the optimal power allocation is already available (prices, users, constraints, etc.). 
But this is not the case when actors have to commit for some future slots before all relevant input information is available. 
For example a user can optimize his current consumption based on the present price (e.g.,~\cite{Galus2009EnergyNetworks}), while knowing future price variations would have enabled him to get an even better payoff; similarly, an EV owner informed of the future electricity price but unable to precisely predict its departure time can do no better than minimizing its \emph{expected} electricity cost~\cite{Rad2015UncertainDeparture}. A robust optimization approach dealing with unknown future prices is taken by Conejo, Morales and Baringo in~\cite{Conejo2010DRM}, the objective being to minimize the daily energy cost~\cite{bental2009robust}. 
Other types of unknown information are brought by the clients yet to come, e.g., the quantity and elasticity of their demands.
\emph{Dynamically adapting algorithms} (also called online algorithms) anticipating and adapting to new inputs must hence be defined for such cross-slot optimization. 
We now turn our attention to such approaches developed in the literature. 

%
A simple version of a dynamic algorithm consists in repeatedly applying static algorithms, namely the ones in previous subsections, each time some new information is available. 
This leads to allocations that are optimal if time slots are independent; but in the general case, things are more complicated, and make specifically designed dynamic algorithms necessary. 
Let us borrow an example from Gerding~\textit{et al.}~\cite{Gerding2011MechanismDesign} to illustrate that. 

%
\vspace{1ex}
\hrule
\vspace{1ex}
\begin{example}\label{ex:carol_david}
Consider two EV clients: Carol's EV is going to stay plugged-in for 2 time slots, while David leaves at the end of the first time slot. 
Their marginal valuations of one unit of energy are claimed to be $[\$10,\$4]$ for Carol, and $[\$5]$ for David, as shown in Table~\ref{tab:ex:dynamic}. 
These values stand for the maximum amount a user is willing to pay for each unit of energy: Carol would like to pay \$10 or less to buy the first unit and \$4 or less for the second, and one unit for \$5 or less is sufficient for David. 
\begin{table}[htbp]
\caption{A dynamic problem setting} \label{tab:ex:dynamic}
\centering
\begin{tabular}{|c|c|c|}
\hline
& Carol & David \\
\hline
\hline
Plug-in time slots & $T_C = \{1,2\}$ & $T_D = \{1\}$\\
\hline
Marginal willingness-to-pay & $v_C = [10,4]$ & $v_D = [5]$\\
\hline
\end{tabular}
\end{table}
Suppose we have one unit of energy available at each time slot, and that our goal is to maximize social welfare (i.e., the total user valuation for the allocated energy). 
If users only report their current willingness-to-pay but not their intended plug-in duration, treating the problem as static leads to allocating the current unit to the user who values it most. 
For our example, Carol would obtain the first time slot (having the highest valuation), and would have no competitor for the second time slot, hence obtaining it again, for a total user benefit of \$10 + \$4 = \$14. 
But this greedy allocation per slot is not optimal: from Table~\ref{tab:ex:dynamic} we remark that allocating the first unit to David and the second to Carol yields a higher total benefit of \$15 = \$5 + \$10. 
To quantify the loss of value due to limited information, a common measure is the ratio of the objective value reached with the algorithm considered, over the optimum value that could have been reached, had all information been available. 
In our example, this efficiency measure equals 14/15.
\vspace{1ex}
\hrule
\vspace{1ex}
\end{example}

%
As evoked before, a possibility when facing new information is to relaunch the decision search (in a myopic way, in the sense that there is no attempt to account for future incoming information). 
Going back to Example~\ref{ex:carol_david}, this method would achieve an efficiency of $1$ if Carol and David truthfully report their plug-in duration and willingness-to-pay, i.e., reveal all the information in Table~\ref{tab:ex:dynamic}. 
But if a third user Edith, with marginal willingness-to-pay \$6, enters the system at the second time slot and leaves immediately after, that information would trigger an allocation update, giving the first-slot unit to Carol (if there is still a chance to do so) and the second one to Edith. 
This allocation will again need to be adjusted if more information arrives, e.g., saying that there will be two units of energy for sale at the second time slot. 
Such a method should yield higher welfare than repeating static algorithms per-slot without adjusting according to newly revealed information, but still does not guarantee to provide the best decisions from the available information (that includes, e.g., probability distributions for the expected future events). 

%
We will henceforth call \emph{dynamic settings}, situations where decisions must be made over time, and not all future information is available: clients dynamically enter and leave the system, there is uncertainty about the set of feasible decisions in the future~\cite{Nisan2007Algorithmic}, etc.

%
Finding efficient solutions for dynamic problems is already complicated, but things can be worse when facing self-interested actors who may be reluctant to reveal information or could strategically report it, as pointed out before. 
For the dynamic energy allocation problem, Gerding~\textit{et al.}~\cite{Gerding2013Mechanism} design a two-sided auction mechanism in which truthful reports (from the selling and buying parties) can be guaranteed by the mechanism in two specific cases (where sellers are myopic, or each buyer is interested in only one time slot).
Otherwise, relaxing the requirement of truthfulness may lead to higher efficiency, by allowing the actors to strategize~\cite{Gerding2013Mechanism}.

Note that very different models for user preferences are considered in the aforementioned references. We therefore believe there is a strong need to survey the current users' economic interests, as well as the potential users' expectations, to build reasonable models and validate them.

\subsubsection{Mobility-based charging management models}
\label{subsubsec:Geometric}
Let us not forget that the primary function of EVs is transportation; this characteristic makes mobility an unavoidable aspect to consider for charging arrangement schemes; be it by simply considering parking periods, or by covering complex mobility plans of EV owners as is done here. First, we take the EV owner point of view when selecting a charging station, then the charging stations point of view through competition to attract users. 

\paragraph{Charging reservation}
\label{para:Reserve}
For EVs facing several options to get energy, guided reservation can reduce the charging delay~\cite{Qin2011ChargingWaiting,Guo2014Rapid}, and charging cost~\cite{Liu2014Routing,Yang2014Taxi}. 

%
Qin and Zhang~\cite{Qin2011ChargingWaiting} design a mechanism to recommend charging stations to EVs traveling in a transportation network, in order to minimize their overall queueing time before getting recharged.
For each on-road EV, only the stations on the shortest path connecting its current location to its destination can be candidates, so none of them will cause any detour.
Each on-road EV periodically sends a reservation request to all reachable stations on the remaining part of its journey; those stations estimate the waiting time for this EV, and the one with the shortest waiting time estimation is reserved. 
This reservation can be adjusted (through cancellation and re-scheduling) at the next round, to dynamically follow the optimal schedule. 
The authors prove a lower-bound of the waiting time, and simulations show that the proposed distributed algorithm achieves a performance close to that bound.

%
Unlike~\cite{Qin2011ChargingWaiting} where the personal information (location and destination) of each EV is revealed to all the potential stations, Guo \textit{et al.}~\cite{Guo2014Rapid} allow the users to keep these information
; even estimating the total time for charging at a specific station (the sum of driving time, waiting time and charging time)  is performed by each EV. 
The estimation is based on the situation of the EV itself and the information received from the power system control center, the intelligent transportation system center, and each charging station.

%
For an EV owner who is more sensitive to the energy cost than to the time consumed, time-dependent electricity pricing provides an opportunity to trade longer traveling and waiting times for cost saving. Liu, Wu and Long~\cite{Liu2014Routing} schedule the charging jointly with the routing in that context. An algorithm is designed to find the path as well as the charging quantity at each station on it, so that the total electricity cost of the journey is minimized. Particularly for a taxi driver, Yang \textit{et al.}~\cite{Yang2014Taxi} study the optimal charging problem for EV taxis with time-varying service incomes and charging costs. They aim at maximizing the long-term average profit of a driver under the constraint of the SOC (state-of-charge) dynamics of the EV battery. It is assumed that the expected revenue from one service time slot and the expected electricity price vary periodically. Those average values and their variation cycles can be learnt by the taxi driver from past experience. At each idle time slot (no passenger onboard), the taxi driver can decide whether to service or to recharge. The authors provide an algorithm and give a closed-form proof of its viability.

\paragraph{Station competition}
\label{para:Competition}

Charging stations compete for customers through prices~\cite{Garzas2012StationSelection}, and may also try to learn the pattern of customers in order to achieve higher revenues~\cite{Gerding2013Mechanism}. 

Garzas and Granados~\cite{Garzas2012StationSelection} assume that all users (informed with the locations of the stations) first send charging requests to all reachable stations, who then broadcast their prices to the users. Finally, each user chooses the cheapest station among all accessible ones. Competition among stations is an oligopoly game~\cite{Kamien1983Conjectural} on prices, where revenues are proportional to prices and to market shares (the latter decreasing as price increases). The cost for producing energy is assumed to follow a convex function.
Simulation results show that this pricing mechanism provides stations with higher utility than the equilibrium price of the Bertrand oligopoly game. 
Users benefit from the price information, saving maximumly 11.5\% with respect to choosing the nearest station. Note that the energy prepared by a station may be below the demand from the actually arrived customers, but the authors claim that the probability of this occurring is very low since users sent requests to many stations \emph{before} choosing where to recharge, so that stations are likely to over-provision energy. Stations can then use the possibly extra energy to serve customers coming without reservation.

%
The scheme in~\cite{Gerding2013Mechanism} described in Section~\ref{subsubsec:SharingFuture} performs a dispatching of clients to separate stations, more specifically, each client is routed to a station where he is entitled with a unit of energy at a time slot convenient to him (a $<$\emph{station, slot}$>$ pair), through a two-sided auction organized by a central controller. 
EVs can make a reservation by reporting their willingness-to-pay matrix over all possible $<$\emph{station, slot}$>$ pairs to the controller, while each station reports the costs of the units of energy it can provide. Upon receiving the reports from both sides, the central controller finds a $<$\emph{station, slot}$>$ pair maximizing the difference (if positive) between the user's willingness-to-pay of this pair and its cost claimed by the station.

Admitting that predicting EV mobility is hard, historical travel surveys can give statistical insights. Since the results on gasoline cars can be safely transplanted to EVs, data sets can be easily found in~\cite{Hejazi2014Data,Gonder2007Data}. Information on user mobility helps the charging stations to better price their energy and organize reservations. Our literature survey shows a very limited number of analytical results for economic models for EV charging encompassing mobility due to the complexity of the models, but the (numerical) results obtained so far suggest this direction has the potential to yield significant improvements.

\subsubsection{The special case of (unidirectional) regulation service and wind-balancing}
\label{subsubsec:UniRegulation}
Load variation, in the sense of supply-demand balancing, has an effect which is equivalent to generation variation. So maneuverable EV charging can offer regulation, in the same way as generation units in conventional power grids. More precisely, when oversupply (resp., supply shortage) occurs, regulation down (resp., up) can be realized by raising up (resp., reducing) the recharging power of EVs. Of course, this implies that the penetration rate of EVs is sufficiently large for such scenarios to make sense: too few EVs would not provide much service, since their batteries would quickly be filled and/or the demand reduction they could offer would be insufficient.
Note that with respect to the aforementioned scenarios, the purpose is no longer to play with the tradeoff between EVs' valuation for energy and generators' production costs, but to complete the task of opposing frequency deviation and maintaining a satisfactory frequency level. In this case, the commercial reward from providing frequency regulation (the most expensive ancillary service~\cite{Kirby2007ancillary}) is potentially very attractive for EV owners. We devote a separate subsection to unidirectional regulation here, and address bidirectional regulation in Section~\ref{subsec:V2GRegulation}. 

%
Sortomme and El-Sharkawi~\cite{Sortomme2011ChargingStrategies} consider an aggregator using EVs to provide regulation services while recharging their batteries. Every time slot (typically, an hour), the aggregator chooses a preferred charging rate for each EV, the actual charging rate being subject to fluctuations around this value due to regulation. The aggregator revenues stem from EV owners (paying for charging their cars) and from the grid (paying for carrying out regulation services). The aggregator's purpose can be to maximize its profit or to reduce the average unit electricity price of users. For both purposes, the authors highlight the need for efficient optimization for the system to benefit both EVs and the aggregator, since simple heuristics lead to significantly poor performance.

%
Bessa~\textit{et al.}~\cite{Bessa2012Bidding} also consider an aggregator recharging EVs while providing regulation services, and compare the revenue of providing only downward regulation with that of providing both downward and upward regulation. 
The conclusion is that two-sided regulation is economically more attractive when capacity payment (payed for keeping a certain regulation capacity plugged, i.e., standing by for being occasionally called upon) is offered, otherwise the uncertainty of the parameters--day-ahead wholesale prices, regulation price, vehicle mobility--plays greater roles, hence the importance of accurate prediction.

Conceptually, regulation is just another type of allocation problem, where the commodity is not electricity but a share of regulation. So algorithms in Section~\ref{subsubsec:SharingUsers} should also work by replacing energy amounts with power increment or decrement amounts. However:
\begin{enumerate} 
\item The regulation service asks for an immediate response (within seconds) and each cycle lasts for a short duration (a few minutes), which requires the algorithms to converge fast enough.
\item Costs for EVs need to be better understood. EVs are supposed to be energy-centric and price-sensitive--their main purpose is to reach a desirable SOC at minimum cost--, but providing regulation imposes extra costs due to the negative effects that rapid power changes have on batteries. Those effects are not directly reflected in EVs' energy valuation functions. Therefore, to dispatch regulation in the same manner as energy, power fluctuations need to be included into utility functions, together with the price and resulting SOC. We did not find representative models in this category for uni-directional regulation, which leaves room for research.
\item In practice, regulation payment is settled on an hourly basis (much larger than the operating cycle, which is a few minutes), and it is a prerequisite for the regulation provider to set aside a sufficiently large regulation capacity (e.g., 0.1MW) and maintain its reliable connection to the grid for at least one hour. Hence, if EVs cannot commit to stay plugged-in that long, their marginal contributions can not be readily obtained, and payment sharing becomes complicated. The Shapley value~\cite{shapley1953value} can be applied in that case; we encourage further propositions based on revenue-sharing tools rather than resource allocation for the specific context of regulation.
\end{enumerate}
Due to the unpredictability of the regulation signal, the models in Section~\ref{subsubsec:SharingSlots} cannot be directly applied either, since they consist in partitioning a flexible load to track a given known profile. These features of the regulation service, and its high profitability, make it a specific allocation problem worth specific research effort.

%
By varying the charging rate, EVs can also help cope with the intermittency of wind generation, as shown by Leterme~\textit{et al.}~\cite{Leterme2014Wind}: wind farms can declare their next-day production in the day-ahead market, based on predictions for generation and EV availability. 
Then at every time slot (e.g., of duration 15 minutes) of the next day, it is a stochastic optimization problem to decide the charging rate of the EVs, to minimize the current production mismatch plus the expected mismatch for the rest of the day.


%
%
\section{Bidirectional energy trading}
\label{sec:Bidirectional energy trading}

%
Bidirectional energy trading refers to the cases where EVs can not only buy electricity from the grid, but also sell it back thanks to the Vehicle-to-Grid (V2G) technology. 
This provides the grid operator with an economical way to balance demand and supply, relying on EV batteries as storage facilities or energy buffers. 
As evoked previously with unidirectional energy flows, here too the EV penetration must be sufficient, so that the contribution of EV batteries to the storage service be significant at the grid scale.
In order to make the storage providing option attractive to self-interested EV owners, a reasonable portion of the benefit should be shared with them.
One possibility of doing so is through bidirectional real-time pricing, i.e., setting prices for both energy directions. 
If user reactions to price signals follow some predictable patterns, then carefully designed price schemes can help leverage the great storage capacity scattered in individual EV batteries. 

%
This section reviews the control mechanisms for bidirectional energy trading. The first subsection introduces models characterizing behaviors of individual users facing time-varying bidirectional electricity prices; then we turn our attention to schemes where EVs are treated as batteries (intermittently) available to support the grid. 

\subsection{Individual arbitrage}
\label{subsec:Arbitrage}
Bidirectional electricity pricing (i.e., one price for buying energy from the grid and another price for selling it back) offers EVs the opportunity to arbitrage, i.e., to buy electricity when prices are low and then wait for the grid to repurchase it back at higher prices. 
Note that the energy transmission and/or AC/DC conversion losses should then be considered. In order for an EV to get a higher arbitrage revenue, the bidirectional electricity prices play critical roles, together with the mobility of the EV. The literature provides two ways of analysis of this setting.

%
Hutson, Venayagamoorthy and Corzine~\cite{Hutson2008Intelligent} propose an algorithm to carry out energy trading between an EV and the grid, based on hourly market clearing price data from California ISO (CAISO)\footnote{\url{http://www.caiso.com/Pages/default.aspx}}. 
The algorithm uses Binary Particle Swarm Optimization to find most profitable buying and selling times throughout a day from the EV owner point of view, while guaranteeing a State-of-Charge (SOC) above requirements. 
The model assumes that  the market clearing price is known in advance, a very strong assumption.

\begin{figure}[htbp]
   \centering
   \includegraphics[width=8cm]{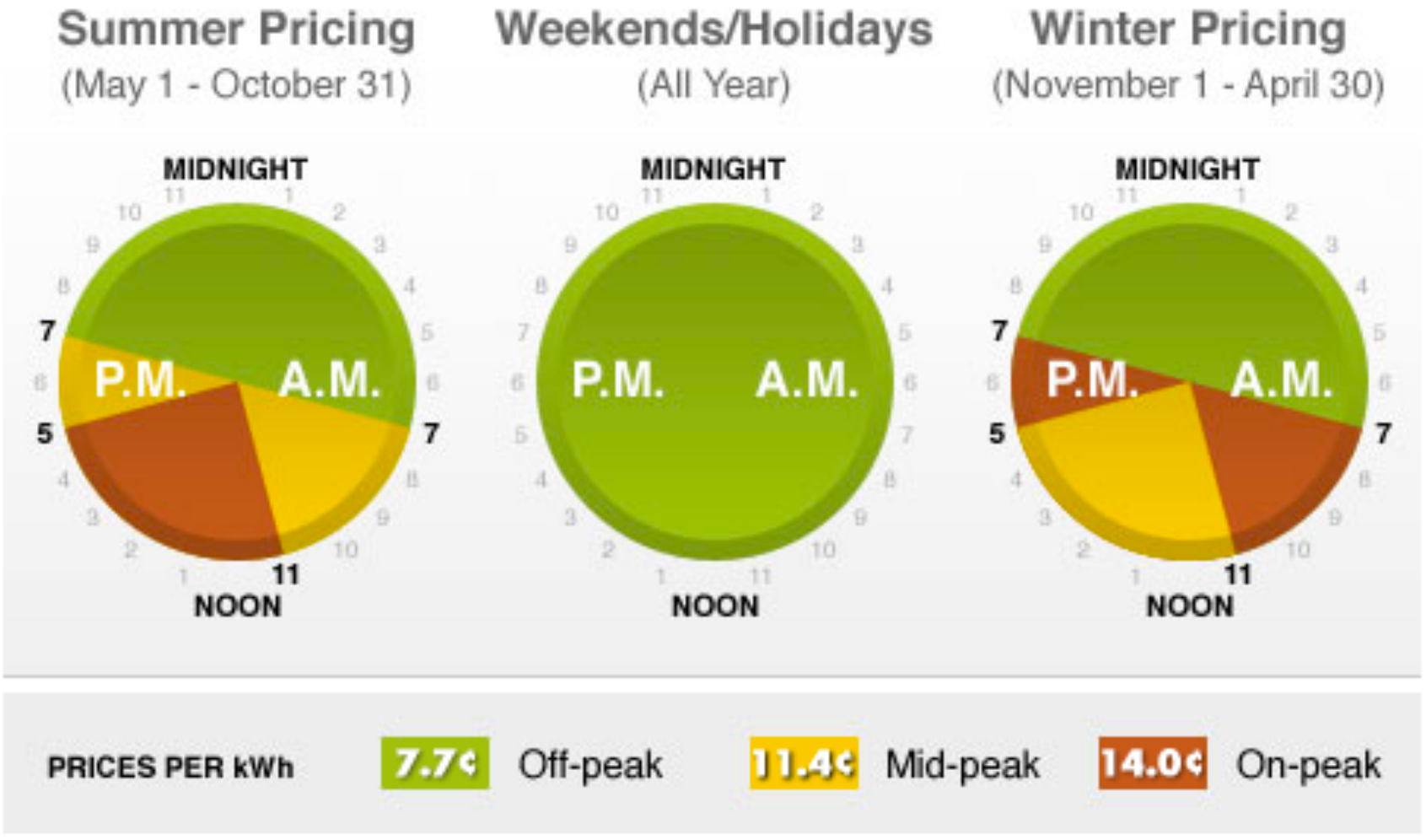} 
   \caption{Ontario Electricity Time-of-use Price periods.}
   \label{fig:TOU-price} 
\end{figure}

%
In the same vein, Liang~\textit{et al.}~\cite{Liang2012V2GStorage} consider a household using a PHEV for daily commute; the householder wants to minimize his energy cost by exchanging electricity with the grid throughout the day, knowing that the electricity price is the Time Of Use (TOU) price in Ontario\footnote{\url{http://www.ontarioenergyboard.ca/OEB/Consumers/Electricity/Electricity+Prices}} as shown in Figure~\ref{fig:TOU-price}. 
The difficulty lies in the (hardly foreseeable) mobility of the user. 
Numerical results indicate that with an estimation of the statistics of the PHEV mobility and energy demand, the proposed scheme performs closely to a scheme with perfect knowledge of the PHEV mobility and energy demand information (efficiency is close to $1$). 
This scheme can then be adjusted when congestion occurs among a group of households, e.g., their aggregated charging (discharging) rate exceeds the upper bound of the power system. 
This high-level adjustment will cause a deviation from the PHEVs' optimal plans, and a cost increase, so the authors further design an adjustment policy such that the power system constraints are satisfied and the incremental cost for PHEVs is minimized~\cite{Liang2013EnergyDelivery}.

\subsection{V2G for regulation services}
\label{subsec:V2GRegulation}

%
Kempton and Letendre~\cite{Kempton1997V2G} proposed the first description of the key concepts of Vehicle to Grid (V2G). 
Their analysis shows that the passenger (combustion) vehicle fleet has ten times the mechanical power of all current American's electrical generation equipment combined, and is idle most of the time. So even with moderate penetration, EVs have the potential to participate in the power market and it is also attractive for the grid operators to let them do so. 
The authors then examine the possibility and profitability of selling EV energy to the grid. 
According to their estimations, the benefit to the grid exceeds the cost to the vehicle owners. 
But this is assuming EVs work as peak power plants, which is not only difficult for them due to their on-board storage limitations~\cite{Kempton2005V2GRevenue}, but also not very financially attractive according to White and Zhang's analysis~\cite{White2011V2G} or even not profitable at all when payment do not compensate the battery degradation~\cite{Zhou2011Battery}. 
So Kempton and Tomi\'c~\cite{Kempton2005V2GRevenue} suggest \emph{regulation} services as a more profitable power market, which better exploits the strengths of EVs: quick response time, low standby costs, and low capital cost per kW. 
A case study of fleets of EVs participating in ancillary services in four US regional regulation markets is provided in~\cite{Kempton2007Fleets}, suggesting that with a few exceptions when the annual market value of regulation was low, V2G power for regulation services is profitable.

%
In the European market, a simulation based on real data, done by Andersson~\textit{et al.} in~\cite{Andersson2010V2GSwedenGermany}, shows that the current German regulating power market would yield significantly higher profits to the PHEVs than the Swedish market. 
They provide a SWOT (Strength, Weaknesses, Opportunities and Threats) analysis of PHEVs as regulating power providers, based on which they portray an ideal regulating power market suited for PHEVs, featured by some key parameters. An ideal regulation market for EVs should provide high capacity payment, allow bidding regulation up and regulation down separately, and have a relatively small bulk bidding size (i.e., 1MW).

%
Considering how scattered and individually owned EVs can participate in the regulation market, Quinn, Zimmerle, and Bradley~\cite{Quinn2010V2G} stress the need for an aggregator, by comparing a centralized architecture (direct communication between EVs and the ISO) with an aggregative (tree-like) architecture---a 3-layer structure involving the ISO, aggregator(s) and EVs. 
The first reason is the relatively low reliability of an individual EV, i.e., the probability of staying plugged-in for a given duration: from 83.6\% to 91.7\% for a time duration of 1 hour, which is incomparable with conventional regulation providers such as natural gas turbines, which have a reliability of 98.89\%~\cite{Quinn2010V2G}. Therefore an aggregator is needed to collect a fleet of EVs so that their reliability be compatible with the current regulation services system requirements. Beside reliability, capacity requirements also call for aggregators: the minimum contractible capacity set by the ISO (from 0.1MW to 10MW in current electricity markets) are indeed way too high for a single EV, due to the battery sizes and the limits of recharging/discharging equipments. 
The aggregator can submit bids to the ISO in the regulation service market, depending on the number (and state) of the EVs it manages. 
During regulation periods, each aggregator then receives a request from the ISO for a certain amount of power (positive or negative) below the contracted regulation capacity. 

Admitting that aggregators are necessary for EVs to be accepted in power markets, the questions arise of how much regulation capacity an aggregator should bid for (normally a bid consists of a capacity and a corresponding price, but we consider only capacity here) to the ISO depending on the number of EVs available and their expected departure times, and how to dispatch the allocated regulation burden among those EVs.
Based on simulations, Kamboj, Decker and Kempton~\cite{Kamboj2010V2G} recommend to dispatch regulation up (down) to EVs whose SOC are above (below) the average level of all. 
The suggested bidding is proportional to the available energy capacity (up and down, in kWh), divided by the regulation duration. 
A scaling parameter quantifying the aggregator degree of conservativeness, is used in the bidding strategy to account for the tradeoff between the revenue and the penalty for not meeting the requested power. 
The authors evaluate this strategy based on real price signal from PJM, the largest transmission operator in the world~\cite{Kamboj2011VandE}, and suggest to share the revenue among EVs according to the Shapley value~\cite{shapley1953value}, a policy with good incentive and fairness properties but computationally difficult to implement. 
The data shows that by providing regulation services for 15 hours a day, an EV can expect to yield one hundred dollars a month of revenues, given the current Regulation Market Clearing Prices.

%
Focusing on regulation dispatch among EVs, Escudero-Garzas, Garcia-Armada and Seco-Granados~\cite{Garzas2012FairDesign} compare several allocation schemes, assuming that the aggregator manages a (sufficiently large) group of EVs available for a known time period (i.e., no mobility is considered). 
Their first scheme maximizes social welfare, that is the total user payoff minus the cost (due to battery degradation), but this may result in a high dispersion among SOCs after regulation. The mechanism is then improved by considering penalties for SOCs approaching the boundaries of some acceptable zones. Maximizing this modified social welfare results in maintaining the variance level among EV SOCs to the one of their arrival time. 
Additionally, the authors suggest a water-filling method (originally used in information theory to maximize the throughput over parallel channels with different channel capacity~\cite{Cover2012IT}): the variance among SOCs keeps decreasing, reaching zero, 
but on the other hand the variance among user payoffs is larger than that after the social welfare maximizing scheme is applied. 
Another aggregator allocation scheme maximizing social welfare is designed by Sun, Dong, and Liang~\cite{Sun2013RegulationAllocation,Sun2014RegulationAllocation}. 
They adopt a general Lyapunov optimization framework and develop a dynamic algorithm to maximize the expected user welfare over an infinite time horizon, which is proven to be asymptotically optimal and performs substantially better than a greedy algorithm optimizing the per-slot system performance.

%
But EVs are not solely regulation providers, they have individual travel plans. 
Specifically, consider an EV who wants to charge itself to a target SOC before a predetermined departure time, at minimum cost. 
Han, Han and Sezaki~\cite{Han2010OptimalV2G} suppose that this user has two choices for each plugged-in hour: recharging, or regulating. 
For the latter, he will be payed a price known in advance for allowing the aggregator to charge or discharge his battery: the uncertainty for the user lies in the direction and amount of the regulation service, out of his control but affecting his outcome.
The proposed solution consists in the user first making a utility-maximizing plan for the whole plugged-in time--assuming null regulation--where utility is the revenue from regulation service minus the charging cost and a punishment based on the discrepancy between the actual SOC on departure and the EV owner's desire. 
Then, since the regulation causes unpredictable (bounded) fluctuations of the SOC, the user relaunches this algorithm again based on the current SOC (hence a static solution to a dynamic problem, as we pointed out in Subsection~\ref{subsubsec:SharingFuture}). 
This method is based on the empirical observation that the time average of regulation requests is almost zero~\cite{Kempton2002V2G}, hence the adjustments from the initial plan remain small. 

On the other hand, the aggregator between the ISO and EVs can be a retailer of regulation services, who first contracts with ISO, then outsources the service to EVs, by setting prices to sell/buy energy to/from EVs to carry out the service; EVs, based on their status and the prices offered by the aggregator, decide whether or not to participate and how much energy to provide or absorb.
Wu, Mohsenian-Rad, and Huang model the relation between the aggregator and EVs as Stackelberg game when providing frequency regulation~\cite{Wu2012V2G} and wind power compensation~\cite{Wu2012Wind}. They design a pricing mechanism to elicit EVs to voluntarily carry out the  services. Among the limitations, let us remark that users in~\cite{Wu2012V2G,Wu2012Wind} are assumed homogenous, i.e., they have identical preferences.
For heterogenous users, a pricing design is provided by Gao~\textit{et al.}~\cite{Gao2013ContractRegulation}: heterogeneity lies in a willingness-to-pay parameter, indicating the users' possibly negative unit value (in monetary unit per kWh) 
for (re)(dis)charging the battery. 
This parameter, compared to the price provided by the aggregator, determines the decision of each EV: upon receiving the regulation power request from the ISO, the aggregator calculates the price so that just enough power from the group of EVs is chosen, taking into account that users are self-interested and rational.
The authors prove the existence of such an optimal price when the distribution of the user parameters follows a regular distribution~\cite{Myerson1978Auction}. 
If the aggregator knows this distribution, it can easily calculate the optimal price and broadcast it to users. Simulations show that the scheme leads to lower prices than~\cite{Wu2012V2G}, hence benefiting the aggregator. When the willingness-to-pay parameter distribution is unknown, the aggregator can implement a learning algorithm to fix the optimal price, using interactions with EVs.

%

\subsection{V2G as storage for renewable energy}
\label{subsec:V2GStorage}

%
Wind farm and solar generation are vagary. 
This plays as a barrier for renewable energy to be widely and efficiently used. 
Indeed, the day-ahead market requires reliable production, and mismatches between submitted bid and real-time injection are sanctioned. 
EVs, with their on-board batteries, can provide storage services through V2G technology, i.e., absorb the surplus and release it when necessary, to maintain a stable output level, or more specifically, to minimize the discrepancy between the real-time output and the day-ahead bidding. 
This can greatly help the development of wind energy according to Kempton and Tomi{\'c}'s calculations~\cite{Kempton2005V2RenewableEnergy}, suggesting that V2G could stabilize large-scale (one-half of US electricity) wind power with 3\% of the fleet dedicated to regulation for wind, plus 8-38\% of the fleet providing operating reserves or storage for wind. 
In terms of expenses, Budischak~\textit{et al.}~\cite{Kempton2013Wind} estimate that the electricity system can be powered 90\% to 99.9\% of the time entirely on renewable electricity, at costs comparable to today's, if we optimize the mix of generation and storage technologies including EV fleets.

\begin{figure}[htbp]
   \centering
   \includegraphics[width=4cm]{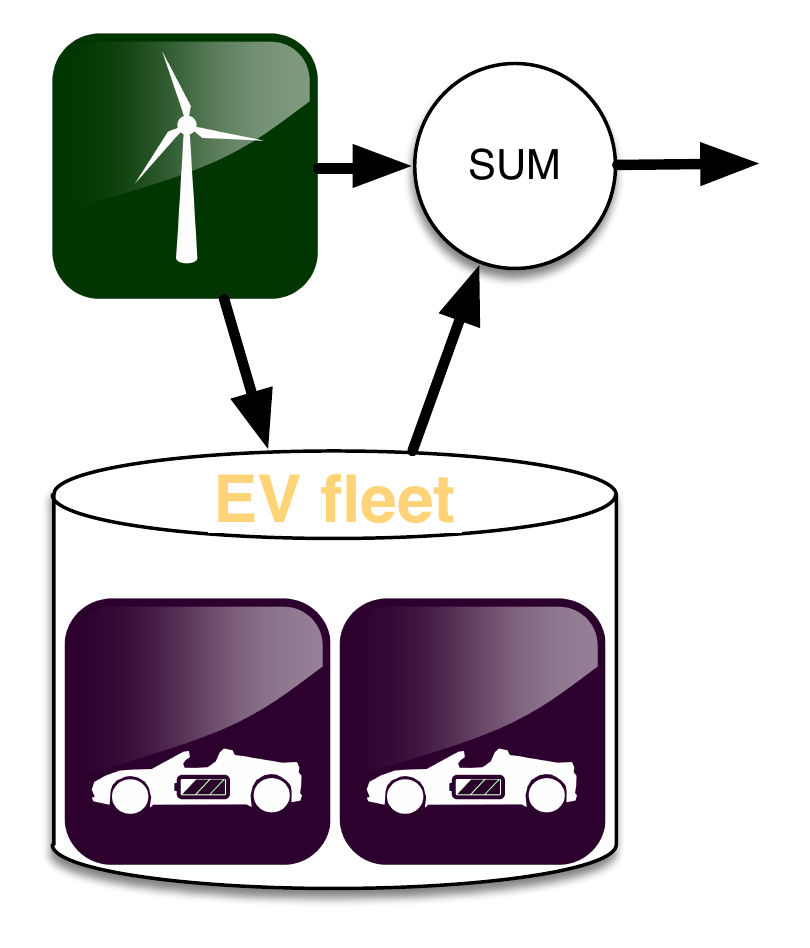} 
   \caption{A virtual power plant, with energy flows.}
   \label{fig:VPP}
\end{figure}

%
To optimize generation and storage, one difficulty lies in providing incentives to attract enough EVs to temperately donate their batteries, and in designing schedules to make the best of them. 
Vasirani~\textit{et al.}~\cite{Vasirani2013VPP} model a Virtual Power Plant (VPP) with EVs providing storage services, as shown in Figure~\ref{fig:VPP}, where the reward to individual EVs is not monetary, but consists in free electricity, proportional to the storage it provides to the VPP.
The VPP bids in the day-ahead market on how much energy it is going to inject every hour for the next day. 
These amounts are based on generation predictions, and take into account the storage system. 
During the next day, the VPP repeatedly searches for the optimal amount of energy to store in (or to withdraw from) EV batteries every hour, as the prediction gets more accurate over time. 
The feasibility of this approach is confirmed through a realistic case-study, using real wind power generation data, corresponding electricity market prices and EVs' characteristics.

%
Xie~\textit{et al.}~\cite{Xie2012Wind} use a similar model to minimize the impact of wind farm production variations. 
They compare two settings: in the first one EVs cooperate with the wind farm by allowing it to use their batteries as a buffer; in the second they just use their batteries to provide frequency regulation to the grid and make revenues, leaving the wind farm undergo the penalties inherent to production variations. 
Numerical results show that the penalty decrease imposed on the wind farm exceeds the decrease of regulation revenue received by the EVs, leaving a negotiation margin to benefit both sides.

\section{Communication aspects}\label{sec:Communication}

Before summarizing the economic properties of the mechanisms proposed in the literature, we first stress the importance of communication systems on their implementation. Information enables decision making and optimization, this section focuses on the content of information exchanges and their frequency. 

\subsection{Information exchanges}
For models that involve two types of actors (i.e., where aggregators and stations are not distinguished and can be referred to as energy vendors), Table~\ref{tab:Info_EV_Agg} presents the main information exchanges that are necessary to implement the schemes described before. 
\begin{table*}[htbp]
\caption{Information exchanges for charging schemes with two types of actors (EV and energy sellers)%
}
\label{tab:Info_EV_Agg}
\begin{center}
\begin{tabular}{cp{.35\textwidth}p{.35\textwidth}}
References&EVs to Energy vendor&Energy vendor to EVs\\

\hline
\cite{Guo2014Rapid}&  &Estimated station availability and charging rate, traffic situation, location\\
\cite{Galus2008DSM}&Willingness-to-pay parameter&Energy allocation\\
\cite{Garzas2012StationSelection}\dag&Charging request (multicast)&Price offers (multicast)\\
\cite{Maille2006Auctions}&Bids (set of pairs \emph{unit price, quantity})&Energy allocation and price\\
\cite{Gerding2011MechanismDesign,Gerding2013Mechanism}&Willingness-to-pay vector or matrix&Energy allocation or reservation\\
\cite{Qin2011ChargingWaiting,Liu2014Routing}&Travel plan, speed, consumption rate, route&Charging plan ( to charge how much energy at which station and at what time)\\

\cite{Samadi2010DSM,Tushar2012Economics,Ardakanian2014DistributedControl}&$\star$Energy demand value &$\star$Price\\
\cite{Beaude2012ChargingGame,Rad2010DSM,Ma2013Valley,Gan2013OptimalProtocol}&$\star$Energy demand vector over discretized time &$\star$Pricing rule, exogenous load or aggregated load of competitors (this forms a penalty while EVs iterate)\\

\rowcolor{grey}
\cite{Han2010OptimalV2G}&Willingness to offer regulation (binary decision)&Capacity and energy prices\\
\rowcolor{grey}
\cite{Gao2013ContractRegulation,Wu2012V2G}&Regulation amount (amount of energy he would like to buy or sell)&Regulation electricity price\\
\rowcolor{grey}
\cite{Sortomme2011ChargingStrategies,Bessa2012Bidding}&Energy demand, charging rate limits, flexibility&Direct control (for regulation)\\
\end{tabular}
\end{center}
\end{table*}
Note that in~\cite{Garzas2012StationSelection} marked with a dagger, the energy vendors are charging stations competing on price to attract EVs; all other models consider a single aggregator as energy vendor, thus research on competition among energy sellers is not abundant. We also highlight that some mechanisms (marked with a star) involve a convergence phase, hence the need for repeated exchanges (with low latency to converge rapidly) before decisions can be made. Grayed cells indicate that regulation services are provided during the charging. References are ranked from the lightest communication burden to the heaviest one.

Some algorithms consider 3 ``layers'' of actors, i.e. EV-Aggregator-ISO or EV-Aggregator-Stations (shown with a double dagger), with the information exchanged shown in Table~\ref{tab:Info_Agg_ISO}. 
The table does not include hardware-related information such as energy transfer efficiency or battery capacity, because they are not crucial for the economic performance of the schemes and often do not need frequent updating, hence have little impact on the communication system. 
Remark also that there can be a tradeoff between communicating and storing: for example in~\cite{Sun2014RegulationAllocation}, the users' accumulating costs can be either sent at every time slot, or recorded with a corresponding user ID. Finally, note that not only information transmission requires communication: so does information retrieval, such as environmental information (wind speed, temperature) that affect energy generation and its forecasting, and user travel record that helps predicting their mobility. 


\begin{table*}[htbp]
\caption{Information exchanges for charging schemes with a 3-layer system (the sequentiality of the exchanges differ among schemes)}
\label{tab:Info_Agg_ISO}
\begin{center}
\begin{tabular}{cp{.2\textwidth}p{.2\textwidth}p{.2\textwidth}p{.2\textwidth}}
References&EVs to Aggregator&Aggregator to ISO (station\ddag)&ISO (station\ddag) to Aggregator&Aggregator to EVs\\
\hline
\cite{Galus2009EnergyNetworks}&Willingness-to-pay parameter& Total energy consumption & Load reduction signal if necessary &Energy allocation \\
\cite{Gerding2013Mechanism}\ddag & Willingness-to-pay vector or matrix & User dispatch & Cost matrix & Energy allocation (reservation of a time slot at a station)\\
\cite{Leterme2014Wind} & SOC & Nothing & Wind generation, and forecasting error probability distribution & Charging power allocation \\
\rowcolor{grey}
\cite{Garzas2012FairDesign} &SOC, acceptable SOC interval, battery cost function & Regulation capacity & Regulation signal (amount and prices)& Regulation allocation\\
\rowcolor{grey}
\cite{Sun2013RegulationAllocation,Sun2014RegulationAllocation} & Utility and cost functions, SOC and accumulated costs & Nothing & Regulation signal (amount) & Regulation allocation\\
\rowcolor{grey}
\cite{Vasirani2013VPP,Xie2012Wind} & SOC cost function & Nothing & generation information, electricity price and/or penalty price &Regulation allocation\\
\end{tabular}
\end{center}
\end{table*}


\subsection{Time granularity}

Table~\ref{Time_Scale} proposes a classification of the approaches presented before, according to the time scale at which they operate. 
Algorithms that update every few seconds are designed for immediate regulation allocation. Regulation requests are sent frequently thus allocations should be computed rapidly. On the other hand, systems reacting to events occurring over time such as supply variations or EV requests can be expected to run less frequently, say, once every few minutes on average. Algorithms running roughly every hour are evoked by the periodic revelation of new environment information such as renewable energy generation or regulation bidding. Long term planning such as day-ahead schedule is made upon precise forecast.

Note that decision updates are driven by new information, so the table also shows the frequency of information exchanges in those algorithms. 

\begin{table*}[htbp]
\caption{Time scale at which charging management operates}
\label{Time_Scale}
\begin{center}
\begin{tabular}{p{4cm}p{4cm}p{4cm}p{4cm}}
Within Seconds	& Within minutes	& Within one Hour	& Day ahead\\
\hline

\cite{Sun2014RegulationAllocation,Wu2012V2G,Gao2013ContractRegulation}(Frequency regulation signal); 

&
~
\cite{Guo2014Rapid} (Supply's variation)
\cite{Garzas2012FairDesign,Sun2013RegulationAllocation} (Regulation settlement every 5 min)\cite{Galus2008DSM,Galus2009EnergyNetworks,Samadi2010DSM,Gerding2013Mechanism,Tushar2012Economics,Maille2006Auctions,Ardakanian2014DistributedControl,Garzas2012StationSelection}(The arrival of supply or EV)

& \tabincell{p{4cm}}{
\cite{Qin2011ChargingWaiting} (Charging reservation updating)\\
\cite{Gerding2011MechanismDesign} (Willingness-to-pay of newly arrived EVs)\\
\cite{Han2010OptimalV2G,Sortomme2011ChargingStrategies,Garzas2012FairDesign} (Hourly settlement of frequency regulation)\\
\cite{Leterme2014Wind,Vasirani2013VPP} (Dynamic forecast of wind generation)\\
}

&
~\cite{Rad2010DSM,Beaude2012ChargingGame,Ma2013Valley,Gan2013OptimalProtocol,Liu2014Routing,Bessa2012Bidding}(A plan for a whole day is made on priori knowledge of price or consumption);
\\
\end{tabular}
\end{center}
\end{table*}

%
%
\section{Classification of approaches and research challenges}
\label{sec:Mechanism}
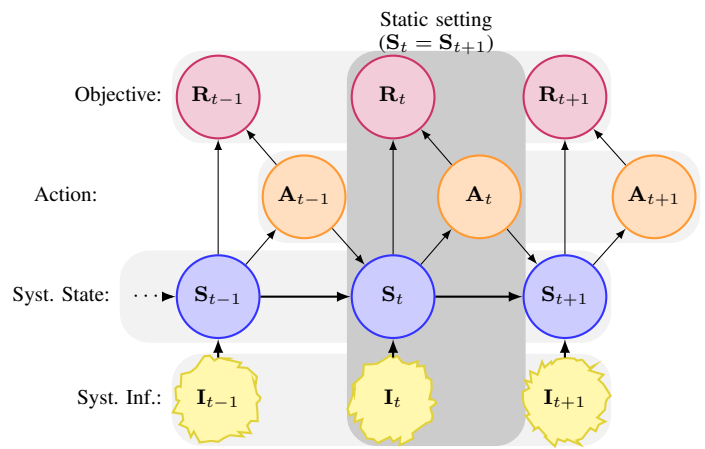
\begin{figure}[htbp]
\centering
\tikzstyle{reward}=[circle,
                                    thick,
                                    minimum size=1.1cm,
                                    draw=purple!80,
                                    fill=purple!20]
\tikzstyle{state}=[circle,
                                    thick,
                                    minimum size=1.1cm,
                                    draw=blue!80,
                                    fill=blue!20
                                    ]
\tikzstyle{action}=[circle,
                                    thick,
                                    minimum size=1.1cm,
                                    draw=orange!80,
                                    fill=orange!25]
\tikzstyle{noise}=[circle,
                                    thick,
                                    minimum size=1.1cm,
                                    draw=yellow!85!black,
                                    fill=yellow!40,
                                    decorate,
                                    decoration={random steps,
                                                            segment length=2pt,
                                                            amplitude=2pt}]
\tikzstyle{background}=[rectangle,
                                     fill=gray!10,
                                     inner sep=0.05cm,
                                     rounded corners=4mm]

\begin{tikzpicture}[>=latex,text height=1ex,text depth=0.25ex]
  
  \matrix[row sep=0.2cm,column sep=0cm] {
    	&
        \node (r_t-1) [reward]{\footnotesize{$\mathbf{R}_{t-1}$}}; &
        &
        \node (r_t)   [reward]{\footnotesize{$\mathbf{R}_t$}};     &
        &
        \node (r_t+1) [reward]{\footnotesize{$\mathbf{R}_{t+1}$}}; &
        \\
        &&\node (a_t-1) [action] {\footnotesize{$\mathbf{A}_{t-1}$}}; &&
        \node (a_t)   [action] {\footnotesize{$\mathbf{A}_t$}};     &&
        \node (a_t+1) [action] {\footnotesize{$\mathbf{A}_{t+1}$}}; &&
        \\
        \node (s_t-2)         {\footnotesize{$\cdots$}};           &
        \node (s_t-1) [state] {\footnotesize{$\mathbf{S}_{t-1}$}}; &&
        \node (s_t)   [state] {\footnotesize{$\mathbf{S}_t$}};     &&
        \node (s_t+1) [state] {\footnotesize{$\mathbf{S}_{t+1}$}}; &
        \\
        &
        \node (n_t-1) [noise] {\footnotesize{$\mathbf{I}_{t-1}$}}; &&
        \node (n_t)   [noise] {\footnotesize{$\mathbf{I}_t$}};&&
        \node (n_t+1) [noise] {\footnotesize{$\mathbf{I}_{t+1}$}};
        ;
        \\
    };
    
    \path[->]
        (s_t-2) edge[thick] (s_t-1)
        (s_t-1) edge[thick] (s_t)
        (s_t)   edge[thick] (s_t+1)
        
        (n_t-1) edge[thick] (s_t-1)
        (n_t) edge[thick] (s_t)
        (n_t+1) edge[thick] (s_t+1)

        (s_t-1) edge (r_t-1)
        (s_t)   edge (r_t)
        (s_t+1) edge (r_t+1)
        
        (s_t-1) edge (a_t-1)
        (s_t)   edge (a_t)
        (s_t+1) edge (a_t+1)
        
        (a_t-1) edge (r_t-1)
        (a_t) edge (r_t)
        (a_t+1) edge (r_t+1)
        
        (a_t-1) edge (s_t)
        (a_t) edge (s_t+1)
        ;
    
    \tikzstyle{backgroundstatic}=[rectangle,
                                     fill=gray!40,
                                     inner sep=0.05cm,
                                     rounded corners=4mm]
    \begin{pgfonlayer}{background}
        \node [background,
                    fit=(r_t-1) (r_t+1),
                    label=left:\footnotesize{Objective:}] {};
        \node [background,
                    fit=(a_t-1) (a_t+1),
                    label=left:\footnotesize{Action:\hspace{2cm}~}] {};
        \node [background,
                    fit=(s_t-2) (s_t+1),
                    label=left:\footnotesize{Syst. State:}] {};
        \node [background,
                    fit=(n_t-1) (n_t+1),
                    label=left:\footnotesize{Syst. Inf.:}] {};
        \node [backgroundstatic,
        		   fit=(r_t) (a_t) (n_t),
		   label=above:\footnotesize{\begin{tabular}{c}Static setting\\($\mathbf{S}_t=\mathbf{S}_{t+1}$)\end{tabular}}] {};
    \end{pgfonlayer}
\end{tikzpicture}\caption{Dynamic problem setting}\label{fig:MDP}
\end{figure}

We summarize in Table~\ref{tab:Aggregator} the economic approaches described in Section~\ref{sec:Unidirectional charging} and Section~\ref{sec:Bidirectional energy trading}.
Firstly the models are classified into two categories, namely \textit{static} and \textit{dynamic} ones, defined in Subsection~\ref{subsubsec:SharingFuture}. 
Static models deal with an isolated time interval in which the performance is determined by actions taken during this time, and optimal actions can be found based on current state information. Contrarily, in a dynamic model where system information varies over time, actions should be updated based on state perturbations caused by such sequential revelations, leading to dynamic optimization methods~\cite{Bertsekas2012dynamic,Puterman2014Markov} as illustrated in Figure~\ref{fig:MDP}. As such, the static setting could be seen as a special case where the state is constant (but still depends on the action taken).

We then distinguish the ways decision-makers interact: \emph{optimization-based approaches} correspond to the cases where one central controller imposes his decisions about allocations and/or prices, and is not influenced by any other actor's actions. 
Ideally, such a central controller has access to all the information needed, thus the management problem reduces to a classical optimization problem: the room for research is therefore
\begin{itemize}
\item for static models, in improving the optimization methods in terms of computational efficiency and/or approximation of the optimum;
\item for dynamic models, in increasing the prediction accuracy and designing algorithms that are robust to unpredictable residuals.
\end{itemize}


In contrast, \emph{game-theoretic approaches} refer to the cases where interactions among several rational actors are considered: even if resources are still dispatched by a central controller, the allocations are affected by other actors' selfish behaviors (e.g., bids sent by EVs). 
Here, static problems already lead to complex models, and even for those approaches, analytical proofs of incentive-compatibility are only valid for some very specific utility functions. While the need seems to be for incentive compatible mechanisms in dynamic settings, designing such schemes is still an open research question in many cases. The difficulty often lies in the evolution of knowledge and beliefs (and thus actions) of actors over time, since the actions taken partly reveal one's private information; analyzing the equilibria of such games is extremely complex.

The last main criterion is related to the implementation type of the schemes: \textit{revelation schemes} imply that actors have to exchange information (such as the willingness-to-pay), and can choose strategically what to reveal, hence the importance of properties such as incentive compatibility. On the opposite, \textit{t\^atonnement schemes} involve a convergence of allocations (and often prices) through iterative methods.

A key aspect in several t\^atonnement-based mechanisms is the convergence of the method: here the limits we found are in the convergence speed (especially in dynamic settings: do prices and allocations have time to converge before the setting changes, say, before another EV arrives?). This is barely addressed in the literature, where in addition convergence is only established for some specific types of utility functions, which need validation.

The classification highlights the need for game-theoretic models in dynamic settings. While it is extremely difficult to design incentive-compatible schemes in dynamic settings, it seems capital to us to develop game-theoretic approaches, even if based on t\^atonnement schemes.

%

\begin{table*}[htbp]
\caption{A classification of economic schemes for EV charging. A diamond mark indicates papers considering PHEVs (which can use fossile fuel) rather than BEVs (which can only use the electric energy stored in their battery).}
\begin{center}
\begin{tabular}{|l|l|l|l|}
\hline
	  	&Optimization-based approaches	&			&Game-theoretic approaches\\
\hline
\multirow{2}{*}{Static}	
		&\multirow{2}{*}{\tabular{p{6cm}}
		\cite{Galus2009EnergyNetworks}\PHEV{}\\
		Heuristic demand curtailment per slot\\
		\cite{Bessa2012Bidding,Franco2015MILP}\cite{Hutson2008Intelligent}\PHEV{}\cite{Yang2014Taxi}\PHEV{}
		Optimization made on prediction of unknown future parameters\\
		\cite{Garzas2012FairDesign}\\
		Fair allocation of regulation per slot\\
		\endtabular}
									&\tabular{l}
									Revelation\\ 
									schemes
									\endtabular
												&\tabular{p{6cm}}
												\cite{Galus2008DSM}\PHEV\\
												Auction based on willingness-to-pay, incentive compatibility assumed\\
												\cite{Maille2006Auctions}
												\\Auction based on willingness-to-pay, incentive compatiblility proved
												\endtabular
												\\
\cline{3-4}&					
									&\tabular{l}
									T\^atonnement\\ 
									schemes
									\endtabular
												&\tabular{p{6cm}}
\cite{Gan2013OptimalProtocol,Ardakanian2014DistributedControl,Samadi2010DSM,Tushar2012Economics,Rad2010DSM,Beaude2012ChargingGame,Hu2014Congestion,Hu2015CongestionNetwork},
\cite{Ma2013Valley}\PHEV{}\\
												Stackelberg game between aggregator and EVs, leader is not omniscient (i.e. unaware of user utility function)\\
												\cite{Garzas2012StationSelection}\\
												Oligopoly game among charging stations\\
												\endtabular
												\\
\hline
\multirow{2}{*}{Dynamic}
		&\multirow{2}{*}{\tabular{p{6cm}}
		\cite{Liang2012V2GStorage}\PHEV,~\cite{Liang2013EnergyDelivery}\PHEV\\
		EV mobility is modeled as a Markov chain\\
		\cite{Sortomme2011ChargingStrategies,Vasirani2013VPP,Xie2012Wind},~\cite{Leterme2014Wind}\PHEV\\
		Forecast accuracies increase as time approaching\\
		\cite{Qin2011ChargingWaiting,Han2010OptimalV2G,Binetti2015SRTG}\\
		Dynamically relaunch a static algorithm\\
		\\
		\endtabular}
									&\tabular{l}
									Revelation\\ 
									schemes
									\endtabular	

												&\tabular{p{6cm}}
												\cite{Gerding2011MechanismDesign}\\
												Incentive compatible for dynamically arriving clients.
												\endtabular 
												\\
\cline{3-4}&
									&\tabular{l}
									T\^atonnement\\ 
									schemes
									\endtabular				
												&\tabular{p{6cm}}
												\cite{Guo2014Rapid}\\
												Dynamically relaunch a static algorithm\\
												\cite{Gerding2013Mechanism,Gao2013ContractRegulation}\\
												Sellers learn users' willingness-to-pay dynamically
												\endtabular
												\\

\hline		
\end{tabular}
\end{center}
\label{tab:Aggregator}
\end{table*}%

\section{Conclusion}
\label{sec:Conclusion}
Electric vehicles, in addition to the prospect of being wholly driven by renewable energy, are not only energy-efficient but also cost-efficient~\cite{Dickerman2010IntroEV}, and emit less greenhouse gas than fossil-fuel based transportation. The main risk they incur comes from the negative impacts they may have on the grid, mostly caused by uncontrolled recharging superimposing on other loads, which exacerbates the grid aging. Coordinating recharging and/or discharging not only alleviates those negative effects, but can also help improve the grid by participating to services such as frequency regulation and energy storage for (intermittent) renewable energy generation. These opportunities can be realized in the Smart Grid realm, so EVs and Smart Grid are mutually reinforcing.  
From the EV owner's point of view, organized recharging and discharging offer the possibility to reduce energy costs or even generate profits.

This paper surveys the charging managements schemes of the literature, with a focus on economy-driven mechanisms. The proposed models, often based on optimization and/or game-theory tools, range from the simple sharing of a given energy amount among several customers (a classical problem) to more complex settings covering aspects such as uncertainty about future events, user mobility constraints, charging station positions, and new grid services like regulation. While some interesting mechanisms have been proposed, and perform well on simulation scenarios, we observed a quite limited amount of analytical results due to the increasing complexity of the settings (large number of actors, specific constraints of distribution networks and EV batteries) and the economic constraints (nonalignment of actors' objectives). Hence we think that further research is needed to better understand the key principles to apply when designing charging management schemes.

The present survey highlights the potential of V2G technology to benefit both EV owners and the grid operator, but also the difficulty of distributing those gains to EV owners to incentivize them to cooperate with the grid operator. 
From the literature review, we witness that management of EV charging processes in smart grids has attracted researchers from diverse domains, and we envision more effort will be devoted to this topic. Several research perspectives are promising from our point of view. Firstly, we consider the trend is pointing at Microgrids~\cite{Soshinskaya2014Microgrids}, which are systems with multiple distributed generators and consumers that can switch between Island mode and connected mode: the presence of EVs is likely to increase the autonomy of such systems. Another research perspective regards the charging management of fleets of EVs, from a fleet owner perspective. For example, with the technology of driverless cars getting matured, driverless taxi fleet may emerge, offering new possibilities for charging (and service providing) management. 

Electric vehicles is an extremely fast-developing field. Technology innovations can reform charging management schemes, for example the roadbed infrastructure would enable charging in motion, which would greatly reduce the reliance on battery capacity and change the understanding (and modeling) of ``plug-in" time.  Economic models for such scenarios are still to be defined.


%

\begin{IEEEbiographynophoto}{Wenjing~Shuai}
received her B.S. degree from Northwestern Polytechnical University, China, in 2008 and the M.S. degree from Xidian University, China, in 2011, both in Telecommunication. She is currently a Ph.D. candidate in Telecom Bretagne, France.
Her research interests include electric vehicle charging management and electricity pricing in Smart Grid.
\end{IEEEbiographynophoto}

\begin{IEEEbiographynophoto}{Patrick~Maill\'e}
graduated from Ecole polytechnique and Telecom ParisTech, France, in 2000 and 2002, respectively. He has been an assistant professor at the Networks, Security, Multimedia department of Telecom Bretagne since 2002, where he obtained his Ph.D. in applied mathematics in 2005, followed by a 6-month visit to Columbia University in 2006. His research interests are on game theory and economic concepts applied to telecommunication ecosystems: resource pricing, routing, consequences of user selfishness on network performance.
\end{IEEEbiographynophoto}


\begin{IEEEbiographynophoto}{Alexander~Pelov}
is an Associate Professor of Computer Networks in the "Networking, Multimedia and Security" department a Telecom Bretagne, France. His research focuses on networking protocols for Machine-to-Machine communications, energy efficiency in wireless networks, and protocols and algorithms for Smart Grid applications, most notably related to Smart Meters, sub-metering and Electrical Vehicles. He received his M.Sc. (2005) from the University of Provence, France and Ph.D. (2009) from the University of Strasbourg, France, both in Computer science. 
\end{IEEEbiographynophoto}





\begin{thebibliography}{10}
\providecommand{\url}[1]{#1}
\csname url@samestyle\endcsname
\providecommand{\newblock}{\relax}
\providecommand{\bibinfo}[2]{#2}
\providecommand{\BIBentrySTDinterwordspacing}{\spaceskip=0pt\relax}
\providecommand{\BIBentryALTinterwordstretchfactor}{4}
\providecommand{\BIBentryALTinterwordspacing}{\spaceskip=\fontdimen2\font plus
\BIBentryALTinterwordstretchfactor\fontdimen3\font minus
  \fontdimen4\font\relax}
\providecommand{\BIBforeignlanguage}[2]{{%
\expandafter\ifx\csname l@#1\endcsname\relax
\typeout{** WARNING: IEEEtran.bst: No hyphenation pattern has been}%
\typeout{** loaded for the language `#1'. Using the pattern for}%
\typeout{** the default language instead.}%
\else
\language=\csname l@#1\endcsname
\fi
#2}}
\providecommand{\BIBdecl}{\relax}
\BIBdecl

\bibitem{Dickerman2010IntroEV}
L.~Dickerman and J.~Harrison, ``A new car, a new grid,'' \emph{IEEE Power
  Energy Mag.}, vol.~8, no.~2, pp. 55--61, 2010.

\bibitem{Balram2012Impact}
P.~Balram, T.~Le~Anh, and L.~Bertling~Tjernberg, ``Effects of plug-in electric
  vehicle charge scheduling on the day-ahead electricity market price,'' in
  \emph{IEEE PES International Conference and Exhibition on Innovative Smart
  Grid Technologies (ISGT Europe)}, Oct. 2012, pp. 1--8.

\bibitem{Foley2013Impact}
A.~Foley, B.~Tyther, P.~Calnan, and B.~{\'O}. Gallach{\'o}ir, ``Impacts of
  electric vehicle charging under electricity market operations,''
  \emph{Applied Energy}, vol. 101, pp. 93--102, 2013.

\bibitem{Green2011ImpactReview}
R.~C. Green, L.~Wang, and M.~Alam, ``The impact of plug-in hybrid electric
  vehicles on distribution networks: A review and outlook,'' \emph{Renewable
  and Sustainable Energy Reviews}, vol.~15, no.~1, pp. 544--553, 2011.

\bibitem{Shafiee2013Impact}
S.~Shafiee, M.~Fotuhi-Firuzabad, and M.~Rastegar, ``Investigating the impacts
  of plug-in hybrid electric vehicles on power distribution systems,''
  \emph{IEEE Trans. Smart Grid}, vol.~4, no.~3, pp. 1351--1360, 2013.

\bibitem{Dow2010Impact}
L.~Dow, M.~Marshall, L.~Xu, J.~Aguero, and H.~Willis, ``A novel approach for
  evaluating the impact of electric vehicles on the power distribution
  system,'' in \emph{IEEE Power and Energy Society General Meeting}, July 2010,
  pp. 1--6.

\bibitem{fudenberg1991game}
D.~Fudenberg and J.~Tirole, \emph{Game Theory}.\hskip 1em plus 0.5em minus
  0.4em\relax MIT Press, Cambridge, Massachusetts, 1991.

\bibitem{Osborne1994course}
M.~J. Osborne and A.~Rubinstein, \emph{A Course in Game Theory}.\hskip 1em plus
  0.5em minus 0.4em\relax MIT Press, 1994.

\bibitem{Fan2013SGCommunications}
Z.~Fan, P.~Kulkarni, S.~Gormus, C.~Efthymiou, G.~Kalogridis, M.~Sooriyabandara,
  Z.~Zhu, S.~Lambotharan, and W.~H. Chin, ``Smart grid communications: Overview
  of research challenges, solutions, and standardization activities,''
  \emph{Commun. Surveys Tuts.}, vol.~15, no.~1, pp. 21--38, First Quarter 2013.

\bibitem{Yan2013SGCommunications}
Y.~Yan, Y.~Qian, H.~Sharif, and D.~Tipper, ``A survey on smart grid
  communication infrastructures: Motivations, requirements and challenges,''
  \emph{Commun. Surveys Tuts.}, vol.~15, no.~1, pp. 5--20, First Quarter 2013.

\bibitem{Alsabaan2013VehicularCommun}
M.~Alsabaan, W.~Alasmary, A.~Albasir, and K.~Naik, ``Vehicular networks for a
  greener environment: A survey,'' \emph{Commun. Surveys Tuts.}, vol.~15,
  no.~3, pp. 1372--1388, Third Quarter 2013.

\bibitem{Yilmaz2012ChargingReview}
M.~Yilmaz and P.~Krein, ``Review of integrated charging methods for plug-in
  electric and hybrid vehicles,'' in \emph{IEEE International Conference on
  Vehicular Electronics and Safety (ICVES)}, 2012, pp. 346--351.

\bibitem{Yilmaz2013Charging}
------, ``Review of battery charger topologies, charging power levels, and
  infrastructure for plug-in electric and hybrid vehicles,'' \emph{IEEE Trans.
  Power Electron.}, vol.~28, no.~5, pp. 2151--2169, May 2013.

\bibitem{Wu2011InductiveCharging}
H.~Wu, A.~Gilchrist, K.~Sealy, P.~Israelsen, and J.~Muhs, ``A review on
  inductive charging for electric vehicles,'' in \emph{IEEE International
  Electric Machines and Drives Conference (IEMDC)}, 2011, pp. 143--147.

\bibitem{Khaligh2012Charging}
A.~Khaligh and S.~Dusmez, ``Comprehensive topological analysis of conductive
  and inductive charging solutions for plug-in electric vehicles,'' \emph{IEEE
  Trans. Veh. Technol.}, vol.~61, no.~8, pp. 3475--3489, Oct. 2012.

\bibitem{Shin2012OLEV}
J.~Shin, B.~Song, S.~Lee, S.~Shin, Y.~Kim, G.~Jung, and S.~Jeon, ``Contactless
  power transfer systems for on-line electric vehicle ({OLEV}),'' in \emph{IEEE
  International Electric Vehicle Conference (IEVC)}, Mar. 2012, pp. 1--4.

\bibitem{Lukic2013Inductive}
S.~Lukic and Z.~Pantic, ``Cutting the cord: Static and dynamic inductive
  wireless charging of electric vehicles,'' \emph{IEEE Electrific. Mag.},
  vol.~1, no.~1, pp. 57--64, Sept. 2013.

\bibitem{Mohrehkesh2011Wireless}
S.~Mohrehkesh and T.~Nadeem, ``Toward a wireless charging for battery electric
  vehicles at traffic intersections,'' in \emph{14th International IEEE
  Conference on Intelligent Transportation Systems (ITSC)}, Oct. 2011, pp.
  113--118.

\bibitem{Lee2010OLEV}
S.~Lee, J.~Huh, C.~Park, N.-S. Choi, G.-H. Cho, and C.-T. Rim, ``On-line
  electric vehicle using inductive power transfer system,'' in \emph{IEEE
  Energy Conversion Congress and Exposition ({ECCE})}, Sept. 2010, pp.
  1598--1601.

\bibitem{DefTSO}
{European Parliament and Council of the European Union}, ``{Directive
  2009/72/EC of the European Parliament and of the Council},'' \emph{Official
  Journal of the European Union}, July 2009.

\bibitem{DefISO}
{Federal Energy Regulatory Commission}, \emph{Docket No. RM99-2-000; Order No.
  2000}, Federal Energy Regulatory Commission, Dec. 1999.

\bibitem{fudenberg1998theory}
D.~Fudenberg, \emph{The theory of learning in games}.\hskip 1em plus 0.5em
  minus 0.4em\relax MIT Press, 1998.

\bibitem{mankiw2014principles}
N.~G. Mankiw, \emph{Principles of Microeconomics}, 7th~ed.\hskip 1em plus 0.5em
  minus 0.4em\relax South-Western College Pub, 2014.

\bibitem{Gkatzikis2013role}
L.~Gkatzikis, I.~Koutsopoulos, and T.~Salonidis, ``The role of aggregators in
  smart grid demand response markets,'' \emph{IEEE Journal on Selected Areas in
  Communications}, vol.~31, no.~7, pp. 1247--1257, 2013.

\bibitem{Galus2008DSM}
M.~D. Galus and G.~Andersson, ``Demand management of grid connected plug-in
  hybrid electric vehicles ({PHEV}),'' in \emph{Proc. of IEEE Energy 2030
  Conference}, 2008, pp. 1--8.

\bibitem{Galus2009EnergyNetworks}
------, ``Integration of plug-in hybrid electric vehicles into energy
  networks,'' in \emph{IEEE Bucharest Power Tech Conference}, June 2009, pp.
  1--8.

\bibitem{Samadi2010DSM}
P.~Samadi, A.-H. Mohsenian-Rad, R.~Schober, V.~Wong, and J.~Jatskevich,
  ``Optimal real-time pricing algorithm based on utility maximization for smart
  grid,'' in \emph{IEEE International Conference on Smart Grid Communications
  (SmartGridComm)}, 2010, pp. 415--420.

\bibitem{Tushar2012Economics}
W.~Tushar, W.~Saad, H.~Poor, and D.~Smith, ``Economics of electric vehicle
  charging: A game theoretic approach,'' \emph{IEEE Trans. Smart Grid}, vol.~3,
  no.~4, pp. 1767--1778, 2012.

\bibitem{Maille2006Auctions}
P.~Maill\'e and B.~Tuffin, ``Pricing the internet with multibid auctions,''
  \emph{IEEE/ACM Trans. Netw.}, vol.~14, no.~5, pp. 992--1004, Oct. 2006.

\bibitem{Ardakanian2014DistributedControl}
O.~Ardakanian, S.~Keshav, and C.~Rosenberg, ``Real-time distributed control for
  smart electric vehicle chargers: From a static to a dynamic study,''
  \emph{IEEE Trans. Smart Grid}, vol.~5, no.~5, pp. 2295--2305, Sept. 2014.

\bibitem{Qin2011ChargingWaiting}
H.~Qin and W.~Zhang, ``Charging scheduling with minimal waiting in a network of
  electric vehicles and charging stations,'' in \emph{Proc. of the 8th ACM
  International Workshop on Vehicular Inter-networking}, 2011, pp. 51--60.

\bibitem{Guo2014Rapid}
Q.~Guo, S.~Xin, H.~Sun, Z.~Li, and B.~Zhang, ``Rapid-charging navigation of
  electric vehicles based on real-time power systems and traffic data,''
  \emph{IEEE Trans. Smart Grid}, vol.~5, no.~4, pp. 1969--1979, July 2014.

\bibitem{Liu2014Routing}
C.~Liu, J.~Wu, and C.~Long, ``Joint charging and routing optimization for
  electric vehicle navigation systems,'' in \emph{International Federation of
  Automatic Control}, Aug. 2014.

\bibitem{Hu2014Congestion}
J.~Hu, S.~You, M.~Lind, and J.~Ostergaard, ``Coordinated charging of electric
  vehicles for congestion prevention in the distribution grid,'' \emph{IEEE
  Trans. Smart Grid}, vol.~5, no.~2, pp. 703--711, Mar. 2014.

\bibitem{Hu2015CongestionNetwork}
J.~Hu, G.~Yang, and H.~Bindner, ``Network constrained transactive control for
  electric vehicles integration,'' in \emph{IEEE Power Energy Society General
  Meeting}, July 2015, pp. 1--5.

\bibitem{Franco2015MILP}
J.~Franco, M.~Rider, and R.~Romero, ``A mixed-integer linear programming model
  for the electric vehicle charging coordination problem in unbalanced
  electrical distribution systems,'' \emph{IEEE Trans. Smart Grid}, vol.~6,
  no.~5, pp. 2200--2210, Sept. 2015.

\bibitem{Beaude2012ChargingGame}
O.~Beaude, S.~Lasaulce, and M.~Hennebel, ``Charging games in networks of
  electrical vehicles,'' in \emph{6th International Conference on Network
  Games, Control and Optimization (NetGCooP)}, 2012, pp. 96--103.

\bibitem{Rad2010DSM}
A.-H. Mohsenian-Rad, V.~Wong, J.~Jatskevich, R.~Schober, and A.~Leon-Garcia,
  ``Autonomous demand-side management based on game-theoretic energy
  consumption scheduling for the future smart grid,'' \emph{IEEE Trans. Smart
  Grid}, vol.~1, no.~3, pp. 320--331, 2010.

\bibitem{Ma2013Valley}
Z.~Ma, D.~Callaway, and I.~Hiskens, ``Decentralized charging control of large
  populations of plug-in electric vehicles,'' \emph{IEEE Trans. Control Syst.
  Technol.}, vol.~21, no.~1, pp. 67--78, Jan. 2013.

\bibitem{Gan2013OptimalProtocol}
L.~Gan, U.~Topcu, and S.~Low, ``Optimal decentralized protocol for electric
  vehicle charging,'' \emph{IEEE Trans. Power Syst.}, vol.~28, no.~2, pp.
  940--951, May 2013.

\bibitem{Sortomme2011ChargingStrategies}
E.~Sortomme and M.~El-Sharkawi, ``Optimal charging strategies for
  unidirectional {Vehicle-to-Grid},'' \emph{IEEE Trans. Smart Grid}, vol.~2,
  no.~1, pp. 131--138, 2011.

\bibitem{Bessa2012Bidding}
R.~Bessa, M.~Matos, F.~Soares, and J.~Lopes, ``Optimized bidding of a {EV}
  aggregation agent in the electricity market,'' \emph{IEEE Trans. Smart Grid},
  vol.~3, no.~1, pp. 443--452, Mar. 2012.

\bibitem{Leterme2014Wind}
W.~Leterme, F.~Ruelens, B.~Claessens, and R.~Belmans, ``A flexible stochastic
  optimization method for wind power balancing with {PHEV}s,'' \emph{IEEE
  Trans. Smart Grid}, vol.~5, no.~3, pp. 1238--1245, May 2014.

\bibitem{Vasirani2013VPP}
M.~Vasirani, R.~Kota, R.~Cavalcante, S.~Ossowski, and N.~Jennings, ``An
  agent-based approach to virtual power plants of wind power generators and
  electric vehicles,'' \emph{IEEE Trans. Smart Grid}, vol.~4, no.~3, pp.
  1314--1322, Sept. 2013.

\bibitem{Xie2012Wind}
L.~Xie, Y.~Gu, A.~Eskandari, and M.~Ehsani, ``Fast {MPC}-based coordination of
  wind power and battery energy storage systems,'' \emph{Journal of Energy
  Engineering}, vol. 138, no.~2, pp. 43--53, 2012.

\bibitem{Binetti2015SRTG}
G.~Binetti, A.~Davoudi, D.~Naso, B.~Turchiano, and F.~Lewis, ``Scalable
  real-time electric vehicles charging with discrete charging rates,''
  \emph{IEEE Trans. Smart Grid}, vol.~6, no.~5, pp. 2211--2220, Sept. 2015.

\bibitem{Conejo2010DRM}
A.~Conejo, J.~Morales, and L.~Baringo, ``Real-time demand response model,''
  \emph{IEEE Trans. Smart Grid}, vol.~1, no.~3, pp. 236--242, 2010.

\bibitem{Hutson2008Intelligent}
C.~Hutson, G.~Venayagamoorthy, and K.~Corzine, ``Intelligent scheduling of
  hybrid and electric vehicle storage capacity in a parking lot for profit
  maximization in grid power transactions,'' in \emph{IEEE Energy 2030
  Conference}, 2008, pp. 1--8.

\bibitem{Liang2012V2GStorage}
H.~Liang, B.~J. Choi, W.~Zhuang, and X.~Shen, ``Towards optimal energy
  store-carry-and-deliver for {PHEV}s via {V2G} system,'' in \emph{Proc. IEEE
  INFOCOM}, 2012, pp. 1674--1682.

\bibitem{Han2010OptimalV2G}
S.~Han, S.~Han, and K.~Sezaki, ``Development of an optimal {Vehicle-to-Grid}
  aggregator for frequency regulation,'' \emph{IEEE Trans. Smart Grid}, vol.~1,
  no.~1, pp. 65--72, 2010.

\bibitem{Liang2013EnergyDelivery}
H.~Liang, B.~J. Choi, W.~Zhuang, and X.~Shen, ``Optimizing the energy delivery
  via {V2G} systems based on stochastic inventory theory,'' \emph{IEEE Trans.
  Smart Grid}, vol.~4, no.~4, pp. 2230--2243, Dec. 2013.

\bibitem{Yang2014Taxi}
Z.~Yang, L.~Sun, M.~Ke, Z.~Shi, and J.~Chen, ``Optimal charging strategy for
  plug-in electric taxi with time-varying profits,'' \emph{IEEE Trans. Smart
  Grid}, vol.~5, no.~6, pp. 2787--2797, Nov. 2014.

\bibitem{Quinn2010V2G}
C.~Quinn, D.~Zimmerle, and T.~H. Bradley, ``The effect of communication
  architecture on the availability, reliability, and economics of plug-in
  hybrid electric vehicle-to-grid ancillary services,'' \emph{J. Power
  Sources}, vol. 195, no.~5, pp. 1500--1509, 2010.

\bibitem{Kamboj2010V2G}
S.~Kamboj, K.~Decker, K.~Trnka, N.~Pearre, C.~Kern, and W.~Kempton, ``Exploring
  the formation of electric vehicle coalitions for vehicle-to-grid power
  regulation,'' in \emph{AAMAS workshop on agent technologies for energy
  systems (ATES)}, 2010, pp. 1--8.

\bibitem{Kamboj2011VandE}
S.~Kamboj, W.~Kempton, and K.~S. Decker, ``Deploying power grid-integrated
  electric vehicles as a multi-agent system,'' in \emph{Proc. of The 10th
  International Conference on Autonomous Agents and Multiagent Systems
  (AAMAS)}, 2011, pp. 13--20.

\bibitem{Garzas2012FairDesign}
J.~Escudero-Garzas, A.~Garcia-Armada, and G.~Seco-Granados, ``Fair design of
  plug-in electric vehicles aggregator for {V2G} regulation,'' \emph{IEEE
  Trans. Veh. Technol.}, vol.~61, no.~8, pp. 3406--3419, 2012.

\bibitem{Sun2013RegulationAllocation}
S.~Sun, M.~Dong, and B.~Liang, ``Real-time welfare-maximizing regulation
  allocation in aggregator-{EV}s systems,'' in \emph{IEEE Conference on
  Computer Communications Workshops}, Apr. 2013, pp. 13--18.

\bibitem{Sun2014RegulationAllocation}
------, ``Real-time welfare-maximizing regulation allocation in dynamic
  aggregator-{EVs} system,'' \emph{IEEE Trans. Smart Grid}, vol.~5, no.~3, pp.
  1397--1409, May 2014.

\bibitem{Wu2012V2G}
C.~Wu, H.~Mohsenian-Rad, and J.~Huang, ``Vehicle-to-aggregator interaction
  game,'' \emph{IEEE Trans. Smart Grid}, vol.~3, no.~1, pp. 434--442, 2012.

\bibitem{Wu2012Wind}
------, ``{PEV}-based reactive power compensation for wind {DG} units: A
  {Stackelberg} game approach,'' in \emph{IEEE International Conference on
  Smart Grid Communications (SmartGridComm)}, 2012, pp. 504--509.

\bibitem{Gao2013ContractRegulation}
Y.~Gao, Y.~Chen, C.-Y. Wang, and K.~Liu, ``A contract-based approach for
  ancillary services in {V2G} networks: Optimality and learning,'' in
  \emph{Proc. IEEE INFOCOM}, Apr. 2013, pp. 1151--1159.

\bibitem{Gerding2011MechanismDesign}
E.~H. Gerding, V.~Robu, S.~Stein, D.~C. Parkes, A.~Rogers, and N.~R. Jennings,
  ``Online mechanism design for electric vehicle charging,'' in \emph{Proc. of
  The 10th International Conference on Autonomous Agents and Multiagent Systems
  (AAMAS)}, 2011, pp. 811--818.

\bibitem{Gerding2013Mechanism}
E.~H. Gerding, S.~Stein, V.~Robu, D.~Zhao, and N.~R. Jennings, ``Two-sided
  online markets for electric vehicle charging,'' in \emph{Proc. of the 12th
  International conference on Autonomous Agents and Multiagent Systems
  (AAMAS)}, 2013, pp. 989--996.

\bibitem{Garzas2012StationSelection}
J.~Escudero-Garzas and G.~Seco-Granados, ``Charging station selection
  optimization for plug-in electric vehicles: An oligopolistic game-theoretic
  framework,'' in \emph{IEEE PES Innovative Smart Grid Technologies (ISGT)},
  2012, pp. 1--8.

\bibitem{vickrey1961counterspeculation}
W.~Vickrey, ``Counterspeculation, auctions, and competitive sealed tenders,''
  \emph{Journal of Finance}, vol.~16, no.~1, pp. 8--37, Mar. 1961.

\bibitem{clarke1971multipart}
E.~H. Clarke, ``Multipart pricing of public goods,'' \emph{Public Choice},
  vol.~11, pp. 17--33, 1971.

\bibitem{groves1973incentives}
T.~Groves, ``Incentives in teams,'' \emph{Econometrica}, vol.~41, no.~3, pp.
  617--631, July 1973.

\bibitem{KELLYF1997fair}
F.~Kelly, ``Charging and rate control for elastic traffic,'' \emph{European
  Transactions on Telecommunications}, vol.~8, no.~1, pp. 33--37, 1997.

\bibitem{kelly1998rate}
F.~P. Kelly, A.~K. Maulloo, and D.~K.~H. Tan, ``Rate control in communication
  networks: Shadow prices, proportional fairness and stability,'' \emph{Journal
  of the Operational Research Society}, vol.~49, pp. 237--252, 1998.

\bibitem{monderer1996potential}
D.~Monderer and L.~S. Shapley, ``Potential games,'' \emph{Games and Economic
  Behaviour}, vol.~14, pp. 124--143, 1996.

\bibitem{Rad2015UncertainDeparture}
H.~Mohsenian-Rad and M.~Ghamkhari, ``Optimal charging of electric vehicles with
  uncertain departure times: A closed-form solution,'' \emph{IEEE Trans. Smart
  Grid}, vol.~6, no.~2, pp. 940--942, Mar. 2015.

\bibitem{bental2009robust}
A.~Ben-Tal, L.~El~Ghaoui, and A.~Nemirovski, \emph{Robust Optimization}.\hskip
  1em plus 0.5em minus 0.4em\relax Princeton University Press, 2009.

\bibitem{Nisan2007Algorithmic}
N.~Nisan, T.~Roughgarden, E.~Tardos, and V.~V. Vazirani, \emph{Algorithmic Game
  Theory}.\hskip 1em plus 0.5em minus 0.4em\relax Cambridge University Press,
  2007.

\bibitem{Kamien1983Conjectural}
M.~I. Kamien and N.~L. Schwartz, ``{Conjectural Variations},'' \emph{Canadian
  Journal of Economics}, vol.~16, no.~2, pp. 191--211, May 1983.

\bibitem{Hejazi2014Data}
H.~Akhavan-Hejazi, H.~Mohsenian-Rad, and A.~Nejat, ``Developing a test data set
  for electric vehicle applications in smart grid research,'' in \emph{IEEE
  80th Vehicular Technology Conference (VTC Fall)}, Sept. 2014, pp. 1--6.

\bibitem{Gonder2007Data}
\BIBentryALTinterwordspacing
J.~Gonder, T.~Markel, M.~Thornton, and A.~Simpson, ``Using global positioning
  system travel data to assess real-world energy use of plug-in hybrid electric
  vehicles,'' \emph{Transportation Research Record: Journal of the
  Transportation Research Board}, vol. 2017, pp. 26--32, 2007. [Online].
  Available: \url{http://dx.doi.org/10.3141/2017-04}
\BIBentrySTDinterwordspacing

\bibitem{Kirby2007ancillary}
B.~Kirby, ``Ancillary services: Technical and commercial insights,''
  W\"artsil\"a North America Inc., Tech. Rep., 2007.

\bibitem{shapley1953value}
L.~S. Shapley, ``A value for $n$-person games,'' in \emph{Contributions to the
  Theory of Games, volume II, Annals of Mathematical Studies}, H.~Kuhn and
  A.~Tucker, Eds.\hskip 1em plus 0.5em minus 0.4em\relax Princeton University
  Press, 1953, pp. 307--317.

\bibitem{Kempton1997V2G}
W.~Kempton and S.~E. Letendre, ``Electric vehicles as a new power source for
  electric utilities,'' \emph{Transportation Research Part D: Transport and
  Environment}, vol.~2, no.~3, pp. 157--175, 1997.

\bibitem{Kempton2005V2GRevenue}
W.~Kempton and J.~Tomi{\'c}, ``Vehicle-to-grid power fundamentals: Calculating
  capacity and net revenue,'' \emph{J. Power Sources}, vol. 144, no.~1, pp.
  268--279, 2005.

\bibitem{White2011V2G}
C.~D. White and K.~M. Zhang, ``Using vehicle-to-grid technology for frequency
  regulation and peak-load reduction,'' \emph{J. Power Sources}, vol. 196,
  no.~8, pp. 3972--3980, 2011.

\bibitem{Zhou2011Battery}
C.~Zhou, K.~Qian, M.~Allan, and W.~Zhou, ``Modeling of the cost of {EV} battery
  wear due to {V2G} application in power systems,'' \emph{IEEE Trans. Energy
  Convers.}, vol.~26, no.~4, pp. 1041--1050, Dec. 2011.

\bibitem{Kempton2007Fleets}
J.~Tomi{\'c} and W.~Kempton, ``Using fleets of electric-drive vehicles for grid
  support,'' \emph{J. Power Sources}, vol. 168, no.~2, pp. 459--468, 2007.

\bibitem{Andersson2010V2GSwedenGermany}
S.-L. Andersson, A.~Elofsson, M.~Galus, L.~G{\"o}ransson, S.~Karlsson,
  F.~Johnsson, and G.~Andersson, ``Plug-in hybrid electric vehicles as
  regulating power providers: Case studies of {S}weden and {G}ermany,''
  \emph{Energy Policy}, vol.~38, no.~6, pp. 2751--2762, 2010.

\bibitem{Cover2012IT}
T.~M. Cover and J.~A. Thomas, \emph{Elements of information theory}.\hskip 1em
  plus 0.5em minus 0.4em\relax John Wiley \& Sons, 2012.

\bibitem{Kempton2002V2G}
S.~Letendre and W.~Kempton, ``The {V2G} concept: A new model for power?''
  \emph{Public Utilities Fortnightly}, vol. 140, no.~4, pp. 16--26, Feb. 2002.

\bibitem{Myerson1978Auction}
R.~B. Myerson, ``\BIBforeignlanguage{ru}{Optimal auction design},''
  \emph{\BIBforeignlanguage{ru}{Mathematics of Operations Research}}, vol.~6,
  no.~1, pp. 58--73, 1978.

\bibitem{Kempton2005V2RenewableEnergy}
W.~Kempton and J.~Tomi{\'c}, ``Vehicle-to-grid power implementation: From
  stabilizing the grid to supporting large-scale renewable energy,'' \emph{J.
  Power Sources}, vol. 144, no.~1, pp. 280--294, 2005.

\bibitem{Kempton2013Wind}
C.~Budischak, D.~Sewell, H.~Thomson, L.~Mach, D.~E. Veron, and W.~Kempton,
  ``Cost-minimized combinations of wind power, solar power and electrochemical
  storage, powering the grid up to 99.9\% of the time,'' \emph{J. Power
  Sources}, vol. 225, pp. 60--74, 2013.

\bibitem{Bertsekas2012dynamic}
D.~P. Bertsekas, \emph{Dynamic Programming and Optimal Control}, 4th~ed.\hskip
  1em plus 0.5em minus 0.4em\relax Athena Scientific, 2012.

\bibitem{Puterman2014Markov}
M.~Puterman, \emph{Markov Decision Processes: discrete stochastic dynamic
  programming}.\hskip 1em plus 0.5em minus 0.4em\relax John Wiley \& Sons,
  2014.

\bibitem{Soshinskaya2014Microgrids}
M.~Soshinskaya, W.~H. Crijns-Graus, J.~M. Guerrero, and J.~C. Vasquez,
  ``Microgrids: Experiences, barriers and success factors,'' \emph{Renewable
  and Sustainable Energy Reviews}, vol.~40, pp. 659--672, 2014.

\end{thebibliography}
\end{document}